\documentclass[showpacs,10pt,aps,pra,reprint,onecolumn,notitlepage,superscriptaddress]{revtex4-1}
\usepackage{amsmath}
\usepackage{bbm}
\usepackage{amsfonts}
\usepackage{amssymb}
\usepackage{bm}
\usepackage{tensor}
\usepackage{graphicx}
\usepackage[]{subfigure}
\usepackage{color}
\graphicspath{{/maybehome/rthompson/RTT2GB_local/Physics/Research/TransOptics/LatexFiles/FRWcloak/Figures/}}

\providecommand{\ed}{\mathrm{d}}
\providecommand{\ma}{\mathbf{A}}
\providecommand{\mf}{\mathbf{F}}
\providecommand{\mg}{\mathbf{G}}
\providecommand{\mJ}{\mathbf{J}}

\providecommand{\mc}{\boldsymbol{\chi}}
\begin{document}

\title{Shrinking cloaks in expanding spacetimes: the role of coordinates and the meaning of transformations in Transformation Optics}

\author{Robert T. Thompson}
\email{robert@cosmos.phy.tufts.edu}
\affiliation{Department of Mathematics and Statistics,
University of Otago,
P.O.\ Box 56, 
Dunedin, 9054,  New Zealand
}
\author{Mohsen Fathi}
\email{mohsen.fathi@gmail.com}
\affiliation{Department of Physics, 
Payame Noor University (PNU),
P.O.\ Box 19395-3697,
Tehran, Iran}

\begin{abstract}
The fully covariant formulation of transformation optics is used to find the configuration of a cloaking device operating in an expanding universe
modelled by a Friedmann-Lema\^{\i}tre-Robertson-Walker spacetime. 
This spacetime cloak is used as a platform for probing the covariant formulation of transformation optics, thereby rigorously enhancing the conceptual understanding of the theory.
By studying the problem in both comoving and physical coordinates we explicitly demonstrate the preservation of general covariance of electrodynamics under the transformation optics procedure.
This platform also enables a detailed study of the various transformations that arise in transformation optics.  
We define a corporeal transformation as the ``transformation'' of transformation optics, and distinguish it from coordinate and frame transformations. 
We find that corporeal transformations considered in the literature have generally been restricted to a subset of all possible corporeal transformations, providing a potential mechanism for increased functionality of transformation optics. 
\\\\
{\textit{keywords}}: transformation optics, cloaking, metamaterials, curved spacetime, covariant electrodynamics

\end{abstract}

\pacs{41.20.-q, 42.70.-a, 04.20.-q, 02.40.Ky}
\maketitle

\section{Introduction}

The prediction  and subsequent laboratory demonstration of an electromagnetic cloaking device has created a fundamentally new approach to the design of electromagnetic and optical devices, called ``transformation optics'' or ``transformation electrodynamics'' (which we shall generically abbreviate as TO)\cite{Pendry2006sc,Leonhardt2006sc,Cummer2006pre,Schurig2006sc}.
The approach has been widely explored over the last decade both theoretically and experimentally, with studies of cloaks of various shapes, sizes, and operating frequencies; and applications beyond cloaking to myriad different optical systems including lenses, waveguides, antennas, and beam splitters.
The transformation method has even been extended beyond electrodynamics to encompass acoustics, elastics, thermodynamics, and even quantum systems.
Several extensive reviews are available which delve into the increasingly voluminous literature related to cloaking and the transformation method, including refs.\ \cite{Leonhardt2009po,Thompson2012aiep}.

The TO design paradigm introduced by Pendry et. al.\ \cite{Pendry2006sc} was based on the covariance of Maxwell's equations under coordinate transformations.
In a nutshell, shifting from one coordinate system to another changes the coefficients in Maxwell's equations, but it was found that these transformed coefficients could be absorbed by a redefinition of the permittivity and permeability within the original coordinates.
The redefined material parameters essentially introduce a non-vacuum medium in place the of the coordinate transformation.

Since a coordinate transformation is a diffeomorphism that leaves the system physically invariant, it is unclear from this nutshell description how TO actually works.
Potential misunderstanding stems from the fact there are several different mathematical operations involved, all of which are usually referred to as some kind of  ``transformation''.  
Even more unfortunately, they are all likely to be referred to as ``coordinate transformations,'' and significant confusion exists within the field as to the nature and meaning of these different transformations.

In ref.\ \cite{Thompson2012aiep}, the idea of passive vs.\ active transformations was discussed.  A passive transformation is a true coordinate transformation that leaves the physical object (a vector, say) unchanged while, for example, rotating the coordinate system.
The same vector has different component values relative to the original and rotated coordinate systems, but still represents the same object.
An active transformation, on the other hand, actually changes the physical quantity while leaving the coordinates unchanged -- rotating the actual vector rather than rotating the coordinates.

In TO, the main transformation at work is one that results in the description of a physical object -- a ``transformation medium'' -- whose parthenogenesis from the vacuum must be in some way related to the notion of an active transformation rather than that of a passive coordinate transformation.
The word \textit{corporeal} means ``having a material existence'' or ``consisting of a material object;'' and in the spirit of this word we introduce the terminology \textit{corporeal transformation} to mean a transformation that results in the description of a transformation medium.
Thus the ``transformation'' of transformation optics is a corporeal transformation, rather than a true coordinate transformation.
In addition to corporeal and coordinate transformations, we will find that \textit{frame} transformations also play an important role in TO.

In this paper we study cloaking in a curved spacetime, specifically in an expanding Friedmann-Lema\^{\i}tre-Robertson-Walker (FLRW) spacetime.  
Friedmann-Lema\^{\i}tre-Robertson-Walker has some physical relevance as the spacetime of modern cosmological models describing our expanding universe, although the actual cloak parameters for such a cloak are of limited practical value, and obtaining the FLRW spacetime cloak parameters is not the primary objective of this paper.
Instead, the FLRW cloak serves as a vehicle for investigating the underlying framework of TO, and there are three main results of this investigation:
\begin{enumerate}
 \item We show that general covariance is carried across the corporeal transformation, meaning that an actual coordinate transformation may be applied either before or after the corporeal transformation without changing the physical nature of the transformation medium.  
Although one hopes this should be true by the construction of the general theory used here it is not entirely obvious and not guaranteed.  
In fact it is not possible to rigorously answer this question within the more widely used constructions of transformation optics that interpret the transformation as a coordinate transformation associated with a fictitious curved spacetime.
 \item The distinction between corporeal, coordinate, and frame transformations is clarified through a detailed study of the FLRW cloak.
 The FLRW cloak is the ideal platform for this work because it is the simplest cloak and spacetime for which the distinctions become readily apparent and most easily explicable.
 \item We find that the set of all corporeal transformations may be divided into two classes which we will call ``coordinate corporeal transformations'' and ``frame corporeal transformations.''  
 The frame corporeal transformations present a new avenue for further study in TO and may be an interesting and useful class of corporeal transformations.
\end{enumerate}

The most natural framework for TO is that of general relativity, where light propagates on a 4-dimensional, possibly curved manifold.
The connection between TO and curved spacetime was first hinted at in ref.\ \cite{Leonhardt2006njp1} where the propagation of light through a fictitious curved spacetime was identified with the propagation of light through a medium residing in flat spacetime.
But that approach was both restrictive in it's scope and included an artificial and non-rigorous interpretation of a coordinate transformation of flat spacetime as a curved spacetime.
Later it was shown how TO may be more rigorously formulated in such a way that the spacetime manifold under consideration is always the spacetime manifold where the device will operate, enabling the TO paradigm to be applied in curved spacetime environments \cite{Thompson2011jo2}.
Although for many years general relativity was seen as an esoteric subject with no real-world applications, it is well known that general relativistic corrections are crucial to the operation of global positioning systems, and so may not be ignorable for future near-Earth TO applications.
Such corrections have been studied for TO in the Schwarzschild spacetime of a massive spherical object such as Earth \cite{Thompson2012jo1}.  
There is, therefore, merit in developing the most complete and general theory possible and in exploring its capabilities and limits.
Friedmann-Lema\^{\i}tre-Robertson-Walker spacetime is a nice choice for this study because while it is relatively simple it exhibits rich physics.  

This paper is organized as follows: In Sec.\ \ref{Sec:EandM} we review the completely covariant, 4-dimensional theory of electrodynamics and dielectric media on spacetime manifolds.  
This form of electrodynamics leads naturally to the formulation of TO described in Sec.\ \ref{Sec:TO}.  
In Sec.\ \ref{Sec:FRWcloak} the FLRW spacetime is described, the appropriate cloak transformations are defined, and the covariant theory of TO is employed to determine the cloak parameters. 
The resultant cloak is subjected to coordinate transformations and studied in the flat space of a local observer, demonstrating that the general covariance of electrodynamics is preserved under the operation of TO.  
In Sec.\ \ref{Sec:Discussion} we interpret the results of Sec.\ \ref{Sec:FRWcloak} and suggest how corporeal transformations considered in the literature to date have generally been restricted to a subset of the possible corporeal transformations, providing a potential mechanism for increased functionality of the TO paradigm.  
A very detailed analysis of light propagation in FLRW spacetime is provided in Appendix \ref{App:Parametrization}, while detailed discussions of coordinate transformations and frame transformations are relegated to Appendices \ref{App:CoordTrans}, \ref{App:FrameTrans}, and \ref{App:CoordFrameTrans}.

\section{Covariant Electrodynamics} \label{Sec:EandM}

The standard vectorial representation of Maxwell's equations 
\begin{subequations} \label{Eq:Maxwell3D}
\begin{align} 
\nabla\cdot\vec{B} & = 0, & \nabla\times\vec{E} +\frac{\partial\vec{B}}{\partial t} & = 0,  \label{Eq:MaxwellHomog} \\
\nabla\cdot\vec{D} & = \rho, & \nabla\times\vec{H} - \frac{\partial\vec{D}}{\partial t} & = \vec{j}, \label{Eq:MaxwellInhomog}
\end{align}
\end{subequations}
makes the assumption of flat, Minkowski spacetime.
But it has been shown that transformation electrodynamics can only be fully understood in the context of a theory that can adequately distinguish spacetime from medium \cite{Thompson2011jo1,Thompson2011jo2}, so it is advantageous to work with a formulation of electrodynamics where this distinction can be easily made, even if the spacetime is flat.
Here we review the salient features of a formulation that is manifestly covariant, containing only tensorial objects and the spacetime metric \cite{Thompson2011jo2,Thompson2012aiep}.

We let spacetime consist of a manifold $M$ along with a symmetric metric $\mathbf{g}$ that describes the shape of the spacetime.
Given a curve on the spacetime manifold, the squared infinitesimal distance between two points on the curve is given by the line element
\begin{equation}
 ds^2 = g_{\mu\nu}dx^{\mu}dx^{\nu}
\end{equation}
where $g_{\mu\nu}$ is the matrix representation of $\mathbf{g}$.
We adopt the sign convention $(-+++)$ and we use units where the speed of light $c=1$.
With these conventions the metric of Minkowski spacetime in Cartesian coordinates is
\begin{equation}
 \eta_{\mu\nu} = \begin{pmatrix}
                  -1 & 0 & 0 & 0\\
		   0 & 1 & 0 & 0\\
		   0 & 0 & 1 & 0\\
		   0 & 0 & 0 & 1
                 \end{pmatrix}.
\end{equation}

Although the scalar and vector potentials are often treated as non-physical mathematical tools, in light of their role in quantization of the field we follow Maxwell himself in maintaining their central role, although now combined into the 1-form potential $\ma = A_{\mu}$, where $A^{\mu}=(\phi,\vec{A})$ in Minkowski spacetime.  The field strength tensor $\mf = F_{\mu\nu}$ is obtained by taking the exterior derivative of $\ma$
\begin{equation}
 \mf = \ed\ma, \quad F_{\mu\nu} = \partial_{\mu}A_{\nu} - \partial_{\nu}A_{\mu}
\end{equation}
This is the 4-dimensional generalization of obtaining $\vec{E}$ and $\vec{B}$ as derivatives of the potentials, so it follows that $\mf$ contains the field components.
In the locally flat space of an observer, the components of $\mf$ form an antisymmetric matrix
\begin{equation} \label{Eq:FComponents}
 F_{\mu\nu} = 
 \begin{pmatrix}
  0 & -E_1 & -E_2 & -E_3\\
  E_1 & 0 & B_3 & -B_2\\
  E_2 & -B_3 & 0 & B_1\\
  E_3 & B_2 & -B_1 & 0
 \end{pmatrix}.
\end{equation}
On the other hand, we let the components of the electric displacement $\vec{D}$ and magnetic field $\vec{H}$ be contained in the excitation tensor $\mg$.  In a locally flat space, the components of $\mg$ form an antisymmetric matrix
\begin{equation} \label{Eq:GComponents}
 G_{\mu\nu} = 
 \begin{pmatrix}
  0 & H_1 & H_2 & H_3\\
  -H_1 & 0 & D_3 & -D_2\\
  -H_2 & -D_3 & 0 & D_1\\
  -H_3 & D_2 & -D_1 & 0
 \end{pmatrix}.
\end{equation}

As they stand, Maxwell's Equations (\ref{Eq:Maxwell3D}) are incomplete and require the specification of an independent relationship between the fields before solutions can be found, and the same holds true when the fields are expressed in tensorial form.
In the Minkowski vacuum, that relationship reduces to equating components $H_a=B_a$ and $D_a=E_a$.  
This can be made more precise by employing the Hodge dual $\star$, a well-defined and natural operation on pseudo-Riemannian manifolds endowed with a metric, such that in vacuum
\begin{equation} \label{Eq:VacConstitutive}
 \mg = \star\mf,
\end{equation}
or, in component form
\begin{equation} \label{Eq:StarF}
 G_{\mu\nu} = (\star \mathbf{F})_{\mu\nu} = \frac12 \sqrt{|g|}\,\epsilon_{\mu\nu\alpha\beta}g^{\alpha\gamma}g^{\beta\delta}F_{\gamma\delta}.
\end{equation}

Consider Earth's oceans for a moment.
A dense object dropped into the ocean will sink, experiencing the inexorable pull of gravity regardless of how much water surrounds it.
But gravity is nothing more than a word used to describe the curvature of spacetime, and from the ocean example it is clear that the spacetime surrounds and permeates everything that exists; it cannot be removed nor ignored.
Realizing that dielectric media exists within, and is permeated by, the spacetime, we seek a formulation of Maxwell's equations that clearly distinguishes any spacetime or metric contributions from the medium contributions.
It has been shown \cite{Thompson2011jo2,Thompson2012aiep} that for linear media such a formulation can be obtained through a simple extension of the vacuum constitutive relation Eq.\ (\ref{Eq:VacConstitutive}) to
\begin{equation} \label{Eq:CovariantConstitutive}
 \mg = \star\mc\mf
\end{equation}
where the tensor $\mc$ encodes parameters equivalent to permittivity, permeability, and magnetoelectric couplings.
In component form the constitutive relation is
\begin{equation} \label{Eq:ConstitutiveComponents}
 G_{\mu\nu} = \star\indices{_{\mu\nu}^{\alpha\beta}}\chi\indices{_{\alpha\beta}^{\sigma\rho}}F_{\sigma\rho}.
\end{equation}
The most general $\mc$ has 36 independent components, and we require that $\chi\indices{_{\alpha\beta}^{\sigma\rho}}$ is independently antisymmetric on each pair of indices, $\alpha\leftrightarrow\beta$ and $\sigma\leftrightarrow\rho$.  
Additional symmetry conditions may be imposed based on thermodynamic arguments, in particular that $\chi\indices{_{\alpha\beta\sigma\rho}}$ is symmetric under the pair exchange $\alpha\beta\leftrightarrow \sigma\rho$.
The general linear constitutive relation Eq.\ (\ref{Eq:CovariantConstitutive}) can be made applicable to the vacuum, thereby superseding Eq.\ (\ref{Eq:VacConstitutive}), by requiring
\begin{equation}
 \mc_{vac}\mf = \mf,
\end{equation}
which is sufficient to specify $\mc_{vac}$ uniquely.
Expanding Eq.\ (\ref{Eq:ConstitutiveComponents}), the coefficients may be collected into the form
\begin{align} \label{Eq:CovariantRep}
 D_a & = \left(\varepsilon^c\right)\indices{_a^b}E_b + \left(\tensor[^b]{\gamma}{^c}\right)\indices{_a^b}B_b, & H_a & = \left(\mu^{-1}\right)\indices{_a^b}B_b + \left(\tensor[^e]{\gamma}{^c}\right)\indices{_a^b}E_b.
\end{align}
or in the more traditional form
\begin{align} \label{Eq:TraditionalRep}
 D_a & = \varepsilon\indices{_a^b}E_b + \tensor[^h]{\gamma}{_a^b}H_b, & B_a & = \mu\indices{_a^b}B_b + \tensor[^e]{\gamma}{_a^b}E_b.
\end{align}
The $3\times 3$ matrices $\bar{\bar{\mu}}^{-1}$, $\bar{\bar{\varepsilon}}^c$, etc.\ of the `covariant representation' Eqs.\ (\ref{Eq:CovariantRep}) may be connected to the $3\times 3$ matrices $\bar{\bar{\mu}}$, $\bar{\bar{\varepsilon}}$, etc.\ of the `traditional representation' Eqs.\ (\ref{Eq:TraditionalRep}) by
\begin{equation}  \label{Eq:ConstitutiveShifta}
 \bar{\bar{\mu}} = \left(\bar{\bar{\mu}}^{-1}\right)^{-1},
 \quad \bar{\bar{\varepsilon}} = \bar{\bar{\varepsilon}}^c-\left(\tensor[^b]{\bar{\bar{\gamma}}}{^c}\right)\bar{\bar{\mu}}\left(\tensor[^e]{\bar{\bar{\gamma}}}{^c}\right), 
 \quad \tensor[^h]{\bar{\bar{\gamma}}}{}=\left(\tensor[^b]{\bar{\bar{\gamma}}}{^c}\right)\bar{\bar{\mu}}, 
 \quad \tensor[^e]{\bar{\bar{\gamma}}}{} = -\bar{\bar{\mu}} \left(\tensor[^e]{\bar{\bar{\gamma}}}{^c}\right).
\end{equation}
Finally, the covariant, 4-dimensional formulation of Maxwell's equations may be summarized as
\begin{equation}
 \ed\mf = 0, \quad \ed\mg = \mJ, \quad \mg = \star\mc\mf,
\end{equation}
where $\mJ$ is the free charge and current source, which in what follows we take to be zero.

\section{Transformation Optics} \label{Sec:TO}

A good first step at generalizing transformation optics was to employ the Plebanski equations to identify a fictitious spacetime with a dielectric material in flat spacetime \cite{Plebanski1960pr,Leonhardt2006njp1,Bergamin2008pra}.  
One disadvantage of the Plebanski based approach is that the final transformation medium is restricted to applications in vacuum Minkowski spacetimes.  
Since the Plebanski equations are furthermore not strictly covariant this approach fails to satisfy the desire for a completely general, manifestly covariant theory of TO.  
A strictly covariant theory is necessary for a complete and rigorous foundation, and to fully understand the role of transformations as studied here.
Several authors contributed to the development of such a covariant theory \cite{Teixeira2007motl,Rahm2007pn,Bergamin2008pra,Horsley2011njp}, but here we review the approach initially presented in refs.\ \cite{Thompson2010pra,Thompson2011jo1,Thompson2011jo2}, as modified slightly in ref.\ \cite{Thompson2012aiep}.

\begin{figure}
	\includegraphics[scale=.4]{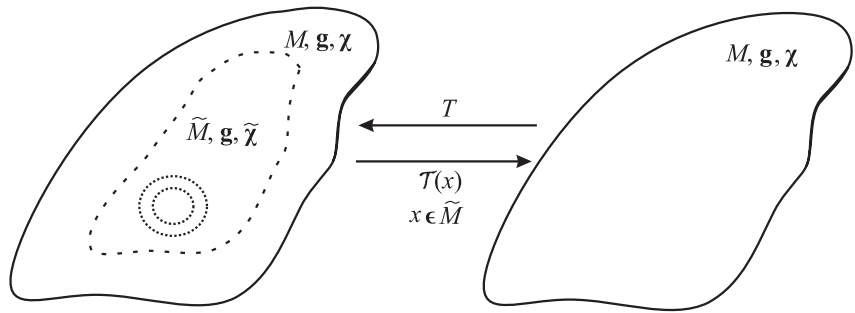}
	\caption{Manifold $M$ is mapped under $T$ to the image $\tilde{M}\subseteq{M}$.  $T$ may be defined as a cloak transformation such that $\tilde{M}$ contains a 'hole' relative to $M$.  However, the fields are transformed by the pullback of $\mathcal{T} = T^{-1}$.}
	\label{Fig:Maps}
\end{figure}

The idea of TO can be formalized as follows.  
Begin with an initial configuration consisting of a spacetime manifold $M$ and metric $\mathbf{g}$, with associated Hodge dual $\star$, and initial material distribution $\mc$ (enabling TO within a prior non-vacuum medium \cite{Thompson2010pra}).  
Let $\mf$ and $\mg$ be the electromagnetic field tensors on $M$, obeying Maxwell's equations and the constitutive relation $\mg = \star\chi\mf$.
Let $T: M\to \tilde{M}\subseteq M$ be a map of $M$ to the image $\tilde{M}$, as in Fig.~\ref{Fig:Maps}.
The image $\tilde{M}$ may not cover all of $M$; for example, $\tilde{M}$ may contain a hole relative to $M$, leading to an invisibility cloak.

Given a map between two manifolds $\varphi:M\to N$ there exists an associated pushforward map $\varphi_*$ that maps vector fields in the tangent space of $M$ to the tangent space of $N$, and a pullback map $\varphi^*$ that maps covector fields from the cotangent space of $N$ to the cotangent space of $M$.
For differential forms like $\mf$ the associated map of interest is the pullback.
Since the pullback map $\varphi^*:T^*N\to T^*M$ is opposite the direction of $\varphi:M\to N$ itself, it follows that in TO we cannot use the pullback of $T:M\to\tilde{M}$ to map $\mf$ from the cotangent space of $M$ to the cotangent space of $\tilde{M}$.  Instead we need the pullback $\mathcal{T}^*$ of a map $\mathcal{T}:\tilde{M}\to M$.  The pullback map $\mathcal{T}^*$ takes the form of the Jacobian matrix of $\mathcal{T}$.

The map $T$ has taken center stage in TO because it provides an intuitive visualization of the desired propagation of light through the transformation medium and an easy method for picking a corporeal transformation based on a mapping of the trajectories of light -- curves on the manifold -- to new trajectories.
In this picture an invisibility cloak can be imagined as mapping the space to a new space with a hole in it, but when calculating the transformation medium we actually require the inverse of $T$; in other words we may take $\mathcal{T} = T^{-1}$.
In fact, TO may be pursued without recourse to the pullback $\mathcal{T}^*$ at all.
Instead, a map $P: T^*M \to T^*\tilde{M}$ between cotangent bundles may be specified independently of a map between manifolds, as further discussed in the following sections.

The essential feature is that the fields $(\mf,\mg)$ are mapped to a new set of fields $\left(\tilde{\mf},\tilde{\mg}\right)$ on $\tilde{M}$, but for these new fields to be valid solutions of Maxwell's equations we require a new `transformation' medium $\tilde{\mc}$.
Furthermore, since TO can not change the spacetime, only the medium, $\tilde{M}\subseteq M$ inherits the metric from $M$.
From the constitutive relations on $M$ and $\tilde{M}$ we find
\begin{equation}
\tilde{\mg} = \mathcal{T}^*(\mg) = \mathcal{T}^*\left(\mc\star\mf\right) = \tilde{\mc}\star\mathcal{T}^*\mf
\end{equation}
which can be solved for the transformation medium \cite{Thompson2010pra,Thompson2011jo1,Thompson2011jo2,Thompson2012aiep}
\begin{equation} \label{Eq:TransOptics}
\tilde{\chi}\indices{_{\eta\tau}^{\pi\theta}}(x)  =  -\star\indices{_{\eta\tau}^{\lambda\kappa}}\Big|_x \Lambda\indices{^{\alpha}_{\lambda}}\Big|_x \Lambda\indices{^{\beta}_{\kappa}}\Big|_x \star\indices{_{\alpha\beta}^{\mu\nu}}\Big|_{\mathcal{T}(x)}\chi\indices{_{\mu\nu}^{\sigma\rho}}\Big|_{\mathcal{T}(x)} (\Lambda^{-1})\indices{^{\pi}_{\sigma}}\Big|_x (\Lambda^{-1})\indices{^{\theta}_{\rho}}\Big|_x.
\end{equation}
In Eq.\ (\ref{Eq:TransOptics}), $\boldsymbol{\Lambda}$ is the Jacobian matrix of $\mathcal{T}$, $\boldsymbol{\Lambda}^{-1}$ is the matrix inverse of $\boldsymbol{\Lambda}$.  The initial material tensor $\mc$ and the first $\star$ are evaluated at $\mathcal{T}(x)$, while everything else is evaluated at $x\in\tilde{M}$.

\section{Cloaking in FLRW Spacetime} \label{Sec:FRWcloak}
\subsection{FLRW spacetime}
In 1929, Edwin Hubble discovered that not only were all distant galaxies receding from us but that the recessional speed of more distant galaxies was faster than the recessional speed of less distant galaxies.
This ph\ae{}nomenon, now referred to as Hubble's Law, appears to be universal and holds true for any choice of origin -- on large scales, everything in the universe is moving away from everything else.
The interpretation of this \textit{Hubble flow} is that the universe is uniformly expanding, with all the galaxies we see going along for the ride.
Not all large-scale motion is due solely to the universal expansion, but as reviewed in Appendix \ref{App:Parametrization}, this extra \textit{peculiar velocity} is damped out as the universe expands, so that eventually all motion will asymptote to the Hubble flow.
Such an expanding universe is well-described by a Friedmann-Lema\^{\i}tre-Robertson-Walker (FLRW) spacetime, which is the general relativistic solution of a spacetime filled with a perfect fluid.

The FLRW spacetime is typically described in ``comoving'' spherical coordinates $(t,r_c,\theta,\varphi)$, with the line element
\begin{equation} \label{Eq:FRWComoving}
ds^2 = -dt^2 + a(t)^2\left(dr_c^2 + r_c^2 d\Omega^2\right)
\end{equation}
and corresponding metric
\begin{equation} \label{Eq:ComovingMetric}
 g_{\mu\nu} = \begin{pmatrix}
  -1 & 0 & 0 & 0 \\
   0 & a^2 & 0 & 0 \\
   0 & 0 & a^2 r_c^2 & 0 \\
   0 & 0 & 0 & a^2 r_c^2 \sin^2\theta
 \end{pmatrix},
\end{equation}
where $a(t)$ is a scale factor that governs the size of a spatial slice of the manifold.  
From the line element one can see that if $a(t)$ is either increasing or decreasing with time then the spatial universe is either expanding or contracting.
For definiteness, in what follows we assume the universe is expanding.
Comoving coordinates are somewhat peculiar in that they are fixed to the spacetime.
Since the spatial universe is expanding, it follows that the physical separation between two points of fixed comoving coordinates is actually increasing with the expansion, similar to the way that the distance between two nearby dots drawn on the surface of a balloon will increase as the balloon is inflated.
An object at rest at radius $r_c$ in comoving coordinates has an increasing physical distance $d = a(t)r_c$.  
Objects at rest in comoving coordinates are said to be comoving, and a comoving object has 4-velocity
\begin{equation} \label{Eq:Comoving4Velocity}
u^{\mu} = (1,0,0,0).
\end{equation}

On the other hand, an observer at the origin would interpret the recession of objects due to universal expansion as some non-zero velocity of the object.
To reflect this observer experience, it is useful to describe FLRW spacetime in ``physical'' coordinates $(t,r_p,\theta,\varphi)$.
Physical coordinates are related to comoving coordinates via the coordinate transformation 
\begin{equation} \label{Eq:ComovingPhysicalTrans}
 r_p = a(t) r_c,
\end{equation}
and in these coordinates the line element becomes
\begin{equation} \label{Eq:FRWPhysical}
ds^2 = -\left(1 - \frac{\dot{a}(t)^2}{a(t)^2} r_p^2\right) dt^2 - 2\frac{\dot{a}(t)}{a(t)}r_p dr_p dt +  dr_p^2 + r_p^2 d\Omega^2
\end{equation}
where an overdot denotes the derivative with respect to $t$, with corresponding metric 
\begin{equation} \label{Eq:PhysicalMetric}
 g_{\mu'\nu'} = \begin{pmatrix}
 -\left(1 - \frac{\dot{a}^2}{a^2} r_p^2\right) & - \frac{\dot{a}}{a}r_p & 0 & 0 \\
  - \frac{\dot{a}}{a}r_p & 1 & 0 & 0 \\
  0 & 0 & r_p^2 & 0 \\
  0 & 0 & 0 & r_p^2\sin^2\theta
 \end{pmatrix}.
\end{equation}
In this and what follows, primed indices refer to physical coordinates while unprimed indices refer to comoving coordinates.
In physical coordinates, a comoving object has 4-velocity
\begin{equation}
u^{\mu'} = (1,\frac{\dot{a}}{a}r_p,0,0),
\end{equation}
with the characteristic feature that the radial velocity depends on the distance, in agreement with Hubble's law and providing compelling evidence that the universe is well-modelled by an expanding FLRW spacetime.
On the other hand, an object at rest in physical coordinates has normalized 4-velocity
\begin{equation} \label{Eq:Physical4Velocity}
u^{\nu'} = \left(\left[1-\left(\frac{ \dot{a}}{a}r_p\right)^2\right]^{-1/2}, 0, 0, 0\right).
\end{equation}

\begin{figure}[ht]
	\subfigure[$ $ Comoving coordinates]{
		\includegraphics[scale=.4]{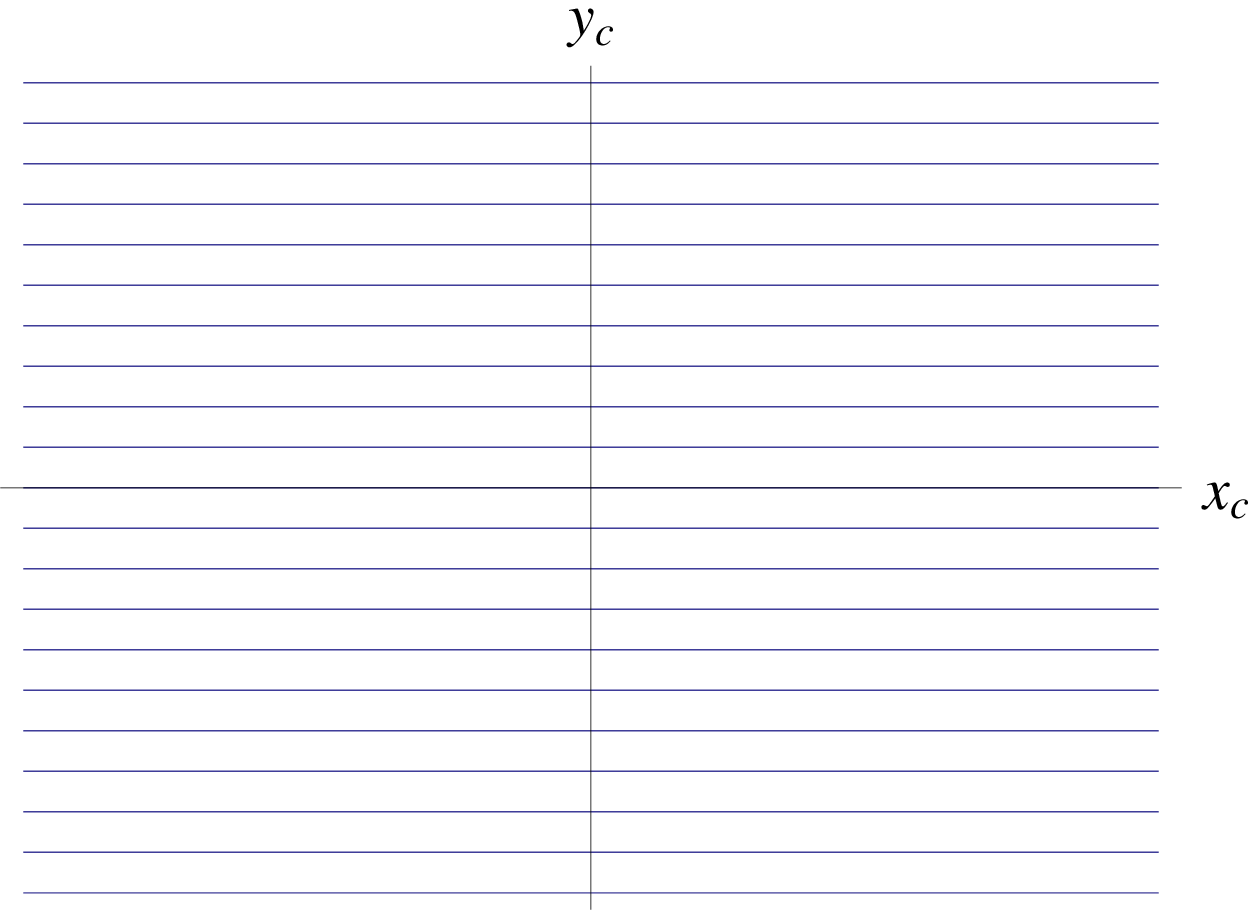}
		\label{Fig:FRWComovingNull}
	}
	\subfigure[$ $ Physical coordinates]{
		\includegraphics[scale=.4]{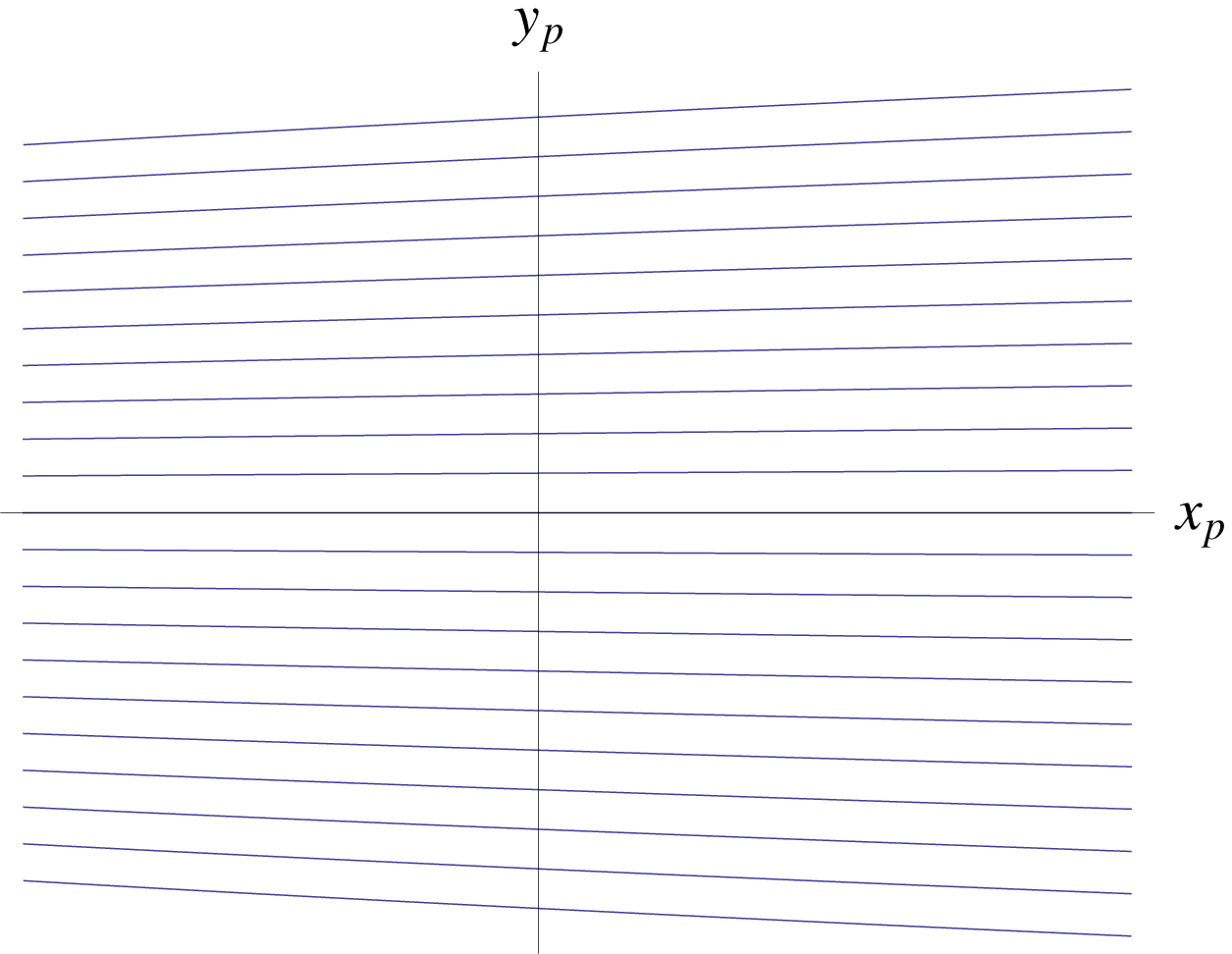}
		\label{Fig:FRWPhysicalNull}
	}
	\caption{Null curves of FLRW spacetime plotted in (a) comoving coordinates, and (b) physical coordinates. Despite the non-zero curvature of FLRW spacetime its conformal flatness is apparent in comoving coordinates, where initially parallel geodesics remain parallel as in Minkowski spacetime.  The expansion of an initially parallel congruence of null geodesics is made clear in physical coordinates.}
	\label{Fig:FRWNullGeodesics}
\end{figure}

The comoving and physical pictures are two equally valid coordinate descriptions of the same spacetime, with a well-defined coordinate transformation to switch between the two. 
Despite the fact that the different coordinates describe the same physics, the geometric optics description of light propagation appears quite different for an observer using comoving vs.\ physical coordinates.
By solving the geodesic equation for null geodesics in each case, as outlined in Appendix \ref{App:Parametrization}, parametric equations of null trajectories are found, with the result plotted in Fig.\ \ref{Fig:FRWNullGeodesics}.
In the figure, the affine parameter is increasing from left to right, and in Fig.\ \ref{Fig:FRWPhysicalNull} the initial conditions specify a congruence of initially 
``parallel'' geodesics.
It is clear from the figure that the notion of parallel does not have its usual meaning in this space of uniform curvature (constant curvature over a constant-time spatial hypersurface).

Note that by making a change of time coordinate, Eq.\ (\ref{Eq:FRWComoving}) can be rewritten as
\begin{equation} \label{Eq:FRWConformal}
 ds^2 = a(\tau)^2\left(-d\tau^2 + dr_c^2 + r_c^2 d\Omega^2\right),
\end{equation}
which is just a conformally scaled version of Minkowski space.
Since conformal maps preserve angles the null geodesics of Minkowski spacetime are conformally mapped to null geodesics in FLRW, thereby preserving the path (although the propagation along the path is different).  Thus it is no surprise that Fig.\ \ref{Fig:FRWComovingNull} looks just like Minkowski spacetime, but we will see that looks can be deceiving when it comes to defining a cloak transformation in the comoving coordinates of FLRW spacetime.

\subsection{Cloak transformations in FLRW}
Given the multiple coordinate choices of FLRW spacetime and the different propagation pictures provided in each, the next step is to choose an appropriate corporeal transformation.
Consider first the Pendry et.\ al.\ transformation \cite{Pendry2006sc} in comoving coordinates
\begin{equation}
r_c \to r_c(1) +\frac{r_c(2)-r_c(1)}{r_c(2)}r_c.
\end{equation}
Bear in mind that this is part of the map $T:M\to \tilde{M}\subseteq M$; the complete specification of $T$ is
\begin{equation}
 T(t,r_c,\theta,\varphi) = (t,r_c(1) +\frac{r_c(2)-r_c(1)}{r_c(2)}r_c, \theta,\varphi).
\end{equation}
Also recall that to calculate the transformation medium $\tilde{\mc}$ with Eq.\ (\ref{Eq:TransOptics}) we actually require the inverse $\mathcal{T} = T^{-1}$ rather than $T$ itself.  

This transformation maps the origin to the cloak's inner surface at fixed radius $r_c(1)$, but in comoving coordinates the fixed radius $r_c(1)$ defines the surface of a physically expanding sphere.
To a local observer, such a comoving cloak would appear to be expanding with the expansion of the universe.
Although one could, with all validity in the context of transformation optics, use this corporeal transformation and calculate the corresponding transformation medium, a physically expanding cloak is not what we had in mind.

Instead, the picture we have in mind is of a cloak of fixed physical size relative to a real, physical observer.  
Every element of a cloak of fixed \textit{physical} size with inner radius $R_1$ and outer radius $R_2$ would actually have some non-zero velocity in comoving coordinates.
This means that the cloak medium would not be comoving, being instead held together by whatever electrostatic forces are present to overcome the universal expansion and keep the cloak from breaking apart.
In other words, a cloak of fixed physical size would actually appear to be shrinking if described in comoving coordinates.

\begin{figure}[ht]
  \includegraphics[scale=.5]{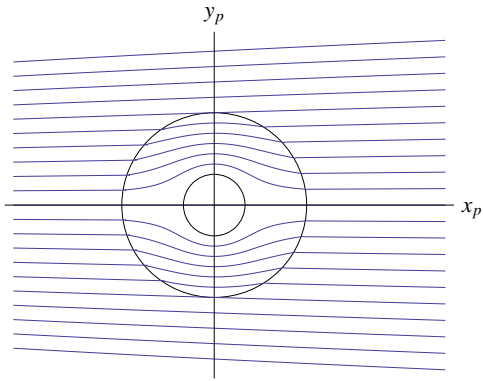}
  \caption{Equatorial slice of a cloak in physical coordinates of FLRW spacetime.  The universal expansion of a set of initially parallel null curves is accommodated by the cloak transformation.}
  \label{Fig:FRWPhysicalCloak}
\end{figure}

On the other hand, in physical coordinates a cloak of fixed physical size would not appear unordinary to a real, physical observer, and there the transformation 
\begin{equation} \label{Eq:PhysicalCloakTrans}
r_p \to R_1 +\frac{R_2-R_1}{R_2}r_p
\end{equation}
would be appropriate.  
It is not immediately obvious whether the Pendry transformation will work in this setting of universal expansion, but the action of a cloak derived from this corporeal transformation in physical coordinates is depicted in Fig.\ \ref{Fig:FRWPhysicalCloak}, and it may be observed that the downstream ray trajectories are perfectly preserved.

Using the coordinate transformation Eq.\ (\ref{Eq:ComovingPhysicalTrans}), the corporeal cloak transformation Eq.\ (\ref{Eq:PhysicalCloakTrans}) may be transformed to comoving coordinates, where the corporeal transformation becomes
\begin{equation} \label{Eq:ComovingCloakTrans}
r_c \to \frac{R_1}{a(t)} +\frac{R_2-R_1}{R_2} r_c
\end{equation}
which for increasing $a(t)$ does indeed represent a shrinking cloak transformation, as predicted for comoving coordinates.

As will be shown, Eqs.\ (\ref{Eq:PhysicalCloakTrans}) and (\ref{Eq:ComovingCloakTrans}) are corporeal transformations for the same cloaking device described by two different coordinates systems.
Recalling that it is the inverse transformation that appears in Eq.\ (\ref{Eq:TransOptics}), we find the Jacobian matrices
\begin{equation} \label{Eq:ComovingJacobian}
 \Lambda\indices{^{\alpha}_{\beta}}(r_c)  = \begin{pmatrix}
  1 & 0 & 0 & 0 \\
  -\frac{R_1R_2 \dot{a}}{(R_2-R_1)a^2} & \frac{R_2}{R_2-R_1} & 0 & 0\\
  0 & 0 & 1 & 0 \\
  0 & 0 & 0 & 1
 \end{pmatrix}
\end{equation}
in comoving coordinates, and
\begin{equation} \label{Eq:PhysicalJacobian}
\Lambda\indices{^{\alpha'}_{\beta'}}(r_p)  = \begin{pmatrix}
1 & 0 & 0 & 0 \\
0 & \frac{R_2}{R_2-R_1} & 0 & 0\\
0 & 0 & 1 & 0 \\
0 & 0 & 0 & 1
\end{pmatrix}
\end{equation}
in physical coordinates.

Using the cloak transformations Eqs.\ (\ref{Eq:ComovingCloakTrans}) and (\ref{Eq:PhysicalCloakTrans}), their associated Jacobian matrices Eqs.\ (\ref{Eq:ComovingJacobian}) and (\ref{Eq:PhysicalJacobian}), and the metrics Eqs.\ (\ref{Eq:ComovingMetric}) and (\ref{Eq:PhysicalMetric}), the tensorial components of the transformation medium $\tilde{\boldsymbol{\chi}}$ may be calculated in each coordinate system through the application of Eq.\ (\ref{Eq:TransOptics}) (we assume the initial space is vacuum, $\mc=\mc_{vac}$).
We write the components as $\chi\indices{_{\alpha\beta}^{\mu\nu}}(x_c)$ and $\chi\indices{_{\alpha'\beta'}^{\mu'\nu'}}(x_p)$, where again unprimed and primed indices refer to comoving and physical coordinates, respectively.

\subsection{Local Frame Results}
However, the coordinate components $\chi\indices{_{\alpha\beta}^{\mu\nu}}(x_c)$ and $\chi\indices{_{\alpha'\beta'}^{\mu'\nu'}}(x_p)$ are of somewhat limited value since they refer to coordinate basis fields of the global coordinates, whereas engineers tasked with building the cloaking device would prefer to work in their locally flat frame.
This is because a local observer measures electric and magnetic fields relative to her local frame, and the permeability and permittivity, as derived from $\mc$ in \cite{Thompson2011jo1}, only have their traditional meaning relative to that local frame.

One of the defining features of a classically curved spacetime is that if you look at only a small portion of the manifold it appears to be Minkowskian, in the same way that the curvature of the Earth is ignorable on the local scale.
By saying that the manifold ``appears locally Minkowskian'' we mean that it is possible, at each point, to define a basis $\{e_A\}$ of the tangent space and related basis $\{\theta^A\}$ of the cotangent space such that the line element relative to this basis is
\begin{equation} \label{Eq:FrameCondition}
 ds^2 = \eta_{AB} \theta^A\theta^B
\end{equation} 
where $\eta_{AB}$ is the Minkowski metric.
This is similar to the way the line element takes on different forms in different coordinates, such as Eqs.\ (\ref{Eq:FRWComoving}), (\ref{Eq:FRWPhysical}), and (\ref{Eq:FRWConformal}).  
However, whereas the tangent and cotangent space bases in Eqs.\ (\ref{Eq:FRWComoving}), (\ref{Eq:FRWPhysical}), and (\ref{Eq:FRWConformal}) follow from the choice of coordinates, the so called ``coordinate bases,'' here the tangent and cotangent bases are changed independently of the coordinates, as further described in Appendices \ref{App:CoordTrans} and \ref{App:FrameTrans}.
The main point is that a basis may be chosen such that relative to the chosen basis the metric becomes Minkowskian even though the spacetime is fundamentally curved, which makes this the natural choice for observers working relative to their locally flat environment.

We have seen that using the corporeal transformation Eqs.\ (\ref{Eq:ComovingCloakTrans}) and (\ref{Eq:PhysicalCloakTrans}) results in transformation media $\chi\indices{_{\alpha\beta}^{\mu\nu}}(x_c)$ and $\chi\indices{_{\alpha'\beta'}^{\mu'\nu'}}(x_p)$. 
As described in Appendix \ref{App:FrameTrans}, the components of $\tilde{\mc}$ in a local frame of comoving coordinates are
\begin{equation} \label{EQ:ChiComovingToLocal}
\chi\indices{_{AB}^{CD}}(x_c) = e\indices{_A^{\alpha}} e\indices{_B^{\beta}} e\indices{^C_{\mu}} e\indices{^D_{\nu}} \chi\indices{_{\alpha\beta}^{\mu\nu}}(x_c)
\end{equation}
while the components of $\tilde{\mc}$ in a local frame of physical coordinates are
\begin{equation} \label{Eq:ChiPhysicalToLocal}
\chi\indices{_{A'B'}^{C'D'}}(x_p) = e\indices{_{A'}^{\alpha'}} e\indices{_{B'}^{\beta'}} e\indices{^{C'}_{\mu'}} e\indices{^{D'}_{\nu'}} \chi\indices{_{\alpha'\beta'}^{\mu'\nu'}}(x_p).
\end{equation}
It still remains to find the frame transformation matrices $e\indices{_A^{\alpha}}$ and $e\indices{_{A'}^{\alpha'}}$.
It may be seen from Eqs.\ (\ref{Eq:FrameCondition}) and (\ref{Eq:MetricFrameTrans}) that the frame transformation matrix can be determined from the condition
\begin{equation}
e\indices{_A^{\alpha}} e\indices{_B^{\beta}} g_{\alpha\beta} = \eta_{AB}
\end{equation}
where 
\begin{equation}
\eta_{AB} = \begin{pmatrix}
   -1 & 0 & 0 & 0 \\
    0 & 1 & 0 & 0 \\
    0 & 0 & r_c^2 & 0 \\
    0 & 0 & 0 & r_c^2\sin^2\theta
\end{pmatrix}
\end{equation}
is the Minkowski metric in spherical comoving coordinates.  
An obviously similar condition holds for $e\indices{_{A'}^{\alpha'}}$ in spherical physical coordinates.
In additional to this condition, we demand that the 4-velocity of an observer at rest in the coordinate basis, e.g.\ Eqs.\ (\ref{Eq:Comoving4Velocity}) and (\ref{Eq:Physical4Velocity}), is frame transformed to the four velocity of an observer at rest in the local frame basis.
This is equivalent to setting $e_0^{\mu} = u^{\mu}$.  One finds
\begin{equation} \label{Eq:ComovingFrameTrans}
e\indices{_A^{\alpha}} = \begin{pmatrix}
 1 & 0 & 0 & 0 \\
 0 & a^{-1} & 0 & 0 \\
 0 & 0 & a^{-1} & 0 \\
 0 & 0 & 0 & a^{-1}
\end{pmatrix}
\end{equation}
and
\begin{equation} \label{Eq:PhysicalFrameTrans}
 e\indices{_{A'}^{\alpha'}} = \begin{pmatrix}
 \frac{1}{\sqrt{1-\left(\frac{r_p \dot{a}}{a}\right)^2}} & 0 & 0 & 0 \\
 -\frac{r_p\dot{a}}{a\sqrt{1-\left(\frac{r_p \dot{a}}{a}\right)^2}} & \sqrt{1-\left(\frac{r_p \dot{a}}{a}\right)^2} & 0 & 0 \\
 0 & 0 & 1 & 0 \\
 0 & 0 & 0 & 1 
 \end{pmatrix}
\end{equation}
in comoving and physical coordinates, respectively, with $ e\indices{^C_{\mu}}$ and $e\indices{^{C'}_{\mu'}}$ found from the  inverse transpose of these matrices.
The permeability and permittivity relative to the local observer may now be calculated as follows: The corporeal transformations in each coordinate systems are given by Eqs.\ (\ref{Eq:ComovingCloakTrans}) and (\ref{Eq:PhysicalCloakTrans}); the Jacobian matrices Eqs.\ (\ref{Eq:ComovingJacobian}) and (\ref{Eq:PhysicalJacobian}) of the corporeal transformations are used in Eq.\ (\ref{Eq:TransOptics}) to calculate $\tilde{\mc}$ relative to the coordinate basis; the frame transformation matrices Eqs.\ (\ref{Eq:ComovingFrameTrans}) and (\ref{Eq:PhysicalFrameTrans}) are used in Eq.\ (\ref{Eq:ChiPhysicalToLocal}) to calculate $\tilde{\mc}$ relative to the locally flat frame; the coefficients of $\tilde{\chi}$ are collected into three dimensional matrices through the identifications of Eq.\ (\ref{Eq:ConstitutiveComponents}) (the component identifications may also be found in ref.\ \cite{Thompson2012aiep}), which are finally put into the traditional representation through Eqs.\ (\ref{Eq:ConstitutiveShifta}).
The result is
\begin{subequations} \label{Eq:ComovingPerm}
	\begin{equation}
	 \bar{\bar{\mu}} = \bar{\bar{\varepsilon}} = \begin{pmatrix}
	 \frac{R_2}{R_2-R_1}\frac{\left(r_c-\frac{R_1}{a}\right)^2}{r_c^2} & 0 & 0 \\
	 0 & \frac{R_2\left(R_2-R_1\right)a^2}{\left(R_2-R_1\right)^2a^2 - R_1^2R_2^2\dot{a}^2} & 0 \\
	 0 & 0 & \frac{R_2\left(R_2-R_1\right)a^2}{\left(R_2-R_1\right)^2a^2 - R_1^2R_2^2\dot{a}^2}
	 \end{pmatrix}
	\end{equation}
	\begin{equation}
	\tensor[^e]{\bar{\bar{\gamma}}}{} = -\tensor[^h]{\bar{\bar{\gamma}}}{} = \begin{pmatrix}
	0 & 0 & 0\\
	0 & 0 & -\frac{R_1R_2^2a \dot{a}}{\left(R_2-R_1\right)^2a^2 - R_1^2R_2^2\dot{a}^2} \\
	0 & \frac{R_1R_2^2a \dot{a}}{\left(R_2-R_1\right)^2a^2 - R_1^2R_2^2\dot{a}^2} & 0
	\end{pmatrix}
 	\end{equation}
\end{subequations}
in a local frame of comoving coordinates, and 
\begin{subequations} \label{Eq:PhysicalPerm}
	\begin{equation}
	\bar{\bar{\mu}} = \bar{\bar{\varepsilon}} = \begin{pmatrix}
	\frac{R_2}{R_2-R_1}\frac{\left(r_p-R_1\right)^2}{r_p^2} & 0 & 0 \\
	0 & \frac{\left(R_2-R_1\right)R_2\left(a^2-r_p^2\dot{a}^2\right)}{\left(R_2-R_1\right)^2 a^2 +R_2^2\left(r_p-R_1\right)^2\dot{a}^2} & 0 \\
	0 & 0 & \frac{\left(R_2-R_1\right)R_2\left(a^2-r_p^2\dot{a}^2\right)}{\left(R_2-R_1\right)^2 a^2 +R_2^2\left(r_p-R_1\right)^2\dot{a}^2}
	\end{pmatrix}
	\end{equation}
	\begin{equation}
	\tensor[^e]{\bar{\bar{\gamma}}}{} = -\tensor[^h]{\bar{\bar{\gamma}}}{} = \begin{pmatrix}
	0 & 0 & 0\\
	0 & 0 & -\frac{R_1\dot{a}}{a}\frac{\left(R_2^2+R_1r_p-2R_2r_p\right)a^2+R_2^2r_p\left(r_p-R_1\right)\dot{a}^2}{\left(R_2-R_1\right)^2 a^2 +R_2^2\left(r_p-R_1\right)^2\dot{a}^2}\\
	0 & \frac{R_1\dot{a}}{a}\frac{\left(R_2^2+R_1r_p-2R_2r_p\right)a^2+R_2^2r_p\left(r_p-R_1\right)\dot{a}^2}{\left(R_2-R_1\right)^2 a^2 +R_2^2\left(r_p-R_1\right)^2\dot{a}^2} & 0
	\end{pmatrix}
	\end{equation}
\end{subequations}
in the local frame of physical coordinates.
It may be readily seen that setting $a(t)=1$ returns us to Minkowski spacetime where the results for both comoving and physical coordinates reduce to the ordinary flat spacetime Pendry cloak.

\subsection{Coordinate Transformations}
Finally, we can now show that the TO procedure and the resulting transformation medium is generally covariant, meaning that knowing the coordinate transformation to switch between comoving and physical coordinates, we can relate the coordinate expressions of $\boldsymbol{\chi}$ by (see Appendix \ref{App:CoordTrans})
\begin{equation}
\chi\indices{_{\alpha'\beta'}^{\mu'\nu'}}(x_p) =  \left. L\indices{^{\alpha}_{\alpha'}}L\indices{^{\beta}_{\beta'}} \chi\indices{_{\alpha\beta}^{\mu\nu}}(x_c) J\indices{^{\mu'}_{\mu}}J\indices{^{\nu'}_{\nu}} \right|_{x_c\to \mathcal{C}(x_c)}
\end{equation}
where $J\indices{^{\mu'}_{\mu}}$ is the Jacobian matrix of the coordinate transformation $x_p = \mathcal{C}(x_c)$ and $L\indices{^{\alpha}_{\alpha'}}$ is the Jacobian matrix of the inverse coordinate transformation $x_c = \mathcal{C}^{-1}(x_p)$.
As applied to a tensor, the coordinate transformation is enacted by contracting the Jacobian matrices of the coordinate transformation and inverse coordinate transformation with the tensor.
However, the Jacobian matrices of the coordinate transformation, by construction, refer to the coordinate basis of the tangent bundle, hence by themselves can only transform from the coordinate basis representation in one coordinate system to the coordinate basis representation in the other coordinate system.
Instead, what we would really like is to transform from the local frame of comoving coordinates to the local frame of physical coordinates.
As shown in Appendix \ref{App:CoordFrameTrans}, the coordinate and frame transformations may be combined such that the components of $\tilde{\mc}$ in one local frame may be directly transformed to the other local frame
\begin{equation} \label{Eq:CoordFrameTrans}
\chi\indices{_{A'B'}^{C'D'}}(x_p) =  K\indices{_{A'}^A} K\indices{_{B'}^B} M\indices{^{C'}_C}M\indices{^{D'}_D}  \left. \chi\indices{_{AB}^{CD}}(x_c)\right|_{x_c\to \mathcal{T}(x_c)}
\end{equation}
with
\begin{equation}
K\indices{_{A'}^A} =  e\indices{^A_{\alpha}} L^{\alpha}_{\alpha'} e\indices{_{A'}^{\alpha'}}
\end{equation}
and
\begin{equation}
M\indices{^{C'}_C} =  e\indices{_C^{\mu}}  J^{\mu'}_{\mu}  e\indices{^{C'}_{\mu'}}.
\end{equation}
It is now possible to verify that Eqs.\ (\ref{Eq:ComovingPerm}) correctly transform to Eqs.\ (\ref{Eq:PhysicalPerm}) under the combined action of the coordinate and frame transformations embodied in Eq.\ (\ref{Eq:CoordFrameTrans}).

There is a widespread misconception within the field of TO that the formulation and results of transformation optics are intimately tied to a choice of coordinates and a subsequent coordinate transformation.
Here we have explicitly demonstrated that this is not the case.
A corporeal transformation given in one coordinate system may be given a different description in another coordinate system, and the resulting transformation media are physically equivalent and are mutually and directly related via a well-defined coordinate transformation.

Although one hopes this to be true by the construction of Eq.\ (\ref{Eq:TransOptics}), it turns out that it is not an entirely trivial or obvious result.
The complete derivation of Eq.\ (\ref{Eq:TransOptics}) requires a more rigorous specification of the metric on each manifold \cite{Thompson2011jo2}; in particular, we may generalize the picture of Fig.\ {\ref{Fig:Maps}} by mapping between different spacetime manifolds.  
This generalization has been used to describe dielectric analog spacetime models that mimic the propagation of light through a vacuum spacetime by the propagation of light through a suitably chosen dielectric medium, and forms the basis for a rigorous derivation of Plebanski's equations \cite{Thompson2010prd}.
However, in that case the coordinates of the target manifold are physically distinct from those of the domain, thereby breaking strict covariance.  
Thus, for example, the FLRW spacetime described in either comoving or physical coordinates may be mapped into an analog medium described by spherical coordinates of Minkowski spacetime, but these will be physically inequivalent representations of the spacetime that cannot be related through a simple coordinate transformation.
The reason covariance is maintained across the corporeal transformation is because in TO we have insisted on the preservation of the underlying spacetime.  
Thus it would seem that formulations of transformation optics based on the Plebanski equations, or equivalently on the analog model, are incapable of rigorously demonstrating the general covariance of the transformation media.
The lack of strict covariance in analog models will be discussed in more detail elsewhere.

\section{Discussion and Conclusions} \label{Sec:Discussion}
We have studied cloaking in an expanding Friedmann-Lema\^{\i}tre-Robertson-Walker spacetime, which is a nice choice because while it is a relatively simple spacetime it is rich in physics and serves as the model for our expanding universe.  
The FLRW cloak makes an ideal platform from which to examine and clarify the role of several different kinds of transformations that may appear in TO. 
In particular it was shown how medium-producing TO transformations, here referred to as corporeal transformations, are distinct from coordinate transformations and frame transformations, and it was shown that all three types of transformations play distinct but important roles in TO.
The corporeal transformation is responsible for what manifests itself as a transformation medium.
Coordinate and frame transformations, on the other hand, simply provide different descriptions of the same physical ph\ae{}nomenon and cannot be used to generate a new medium.
A corporeal transformation may be referred to different coordinate systems, but the physics carried by the corporeal transformation remains invariant under the coordinate transformation, resulting in different coordinate descriptions of the same transformation medium.
In other words the transformation medium obtained from a corporeal transformation in one coordinate system is physically equivalent to the transformation medium obtained when the corporeal transformation is expressed in a different coordinate system, and the two descriptions of the transformation medium are covariantly related by the coordinate transformation.

The FLRW cloak provides a good example of how deciding on the appropriate cloaking transformation depends on the point of view of the observer and their choice of coordinates, but that ultimately the same cloak transformation may be referred to different coordinate systems. 
The desirable scenario of a cloak of fixed physical size was obtained from the usual cloak transformation in physical coordinates, but when referred to comoving coordinates the transformation actually actually describes a radially shrinking cloak.
The cloak transformations in different coordinates are related by the coordinate transformation, and it turns out that this  relationship is preserved across the corporeal map such that the transformation media obtained from either coordinate system are physically equivalent and also related by the coordinate transformation.
Although this coordinate invariance is clearly desirable it is not obvious, and in fact hinges on the fact that the corporeal transformation acts only on the fields, while leaving the manifold untouched. 
This analysis clearly shows that the general covariance of electrodynamics is preserved through the corporeal transformation into the transformation medium, but this need not have been guaranteed.
In fact, the closely related formulation of analog spacetimes \cite{Thompson2010prd} from which the Plebanski equations may be derived and were used in earlier formulations of TO \cite{Leonhardt2006njp1}, breaks covariance under the analog map.

A coordinate transformation acts on a tensor via the differential of the coordinate map.  
Practically, this differential manifests itself as the Jacobian matrix of the transformation, with one index referred to each coordinate system, and is implemented via matrix multiplication; one Jacobian matrix for each tensor index.
The implementation of a frame transformation is also basically matrix multiplication on each tensor index, but while one index of the frame transformation matrix $e\indices{_A^{\mu}}$ refers to the coordinates, the other index refers only to an arbitrary basis vector field that is not, in general, associated to any coordinate system.
As described in Appendix \ref{App:FrameTrans}, a frame field assigns a set of basis vectors at every point, so a frame field is nothing more than a set of four vector fields that may either be chosen freely or chosen to satisfy some conditions, and
any vector field can be decomposed relative to this chosen set of basis vector fields.  Here the frame field is chosen such that the associated metric is Minkowskian.
The arbitrary index on a frame transformation matrix refers to these basis vector fields.

If the frame transformation is smooth then one may find integral curves of the frame basis fields.  If, in addition, the frame fields $\{e_A\}$ commute, $[e_A,e_B]=0$, then the parametrized integral curves may be identified as a set of coordinate functions $\{y^{A}\}$ such that $e_A = \frac{\partial}{\partial y^A}$ and $\theta^A = dy^A$ act as mutually dual bases for the tangent and cotangent bundles.
What this shows is that for commuting frame basis vector fields, the arbitrary index on the frame transformation matrix actually refers to the coordinates corresponding to the integral curves of the fields, and in this case the frame transformation is nothing more than a coordinate transformation.
This provides a refined perspective of the Jacobian matrix, where instead of considering the indices as referring to \textit{coordinates}, they should be thought of as referring to the vector fields whose integral curves correspond to the coordinates, i.e.\ to the coordinate vector fields.
With this refinement, the Jacobian of a coordinate transformation is actually just a special type of frame transformation matrix.

Transformation optics was initially conceived as a coordinate transformation but we have shown that corporeal transformations are distinct from coordinate transformations.  Nevertheless, there is a clear analogy in that each can be identified with a set of functions on the manifold that act upon tensors via the Jacobian matrix of those functions. 
On the other hand, we have seen how coordinate transformations of tensors are a special case of frame transformations; in other words, if $\{\mathfrak{R}\}$ is the set of all possible frame transformation tensors, then the coordinate transformations are the set of all $C\in\mathfrak{R}$ such that $C = d\mathcal{T}$ for some function $\mathcal{T}$.
It follows from the analogy with coordinate transformations that the space of possible corporeal transformations is actually much larger than what has generally been considered in the literature.
In particular, in analogy with a frame transformation, a corporeal transformation may more generally be given through the direct specification of the transformation tensor.
Letting $\{\mathfrak{S}\}$ be the set of all possible corporeal transformation tensors, then the set of all $D\in\mathfrak{S}$ such that $D = d\mathcal{T}$ for some function $\mathcal{T}$ could reasonably be called ``coordinate corporeal transformations.''
Although coordinate transformations are actually a subset of frame transformations, the terminology ``frame transformation'' usually refers to frame transformations that do \textit{not} belong to the subset of coordinate transformations.
By analogy, we may refer to the set of corporeal transformations that are \textit{not} coordinate corporeal transformations as ``frame corporeal transformations.''

With few exceptions \cite{Tretyakov2008njp,Novitsky2012pier}, the multifaceted ``transformation physics'' program has restricted itself to coordinate corporeal transformations and very little seems to have been done with more general classes of transformations.
Frame corporeal transformations therefore present a potential mechanism for increased functionality of the TO paradigm.  
A detailed analysis of these frame corporeal transformations in the context of TO presents a new avenue for further study that lies beyond the scope of the present paper, but is the subject of ongoing investigation.

\appendix
\section{Affine parametrization in spacetime} \label{App:Parametrization}
By applying the map $T$ (see, e.g.\ Fig.\ \ref{Fig:Maps}), a parametrized curve $x^{\mu}(\tau) \subset M$ will be mapped to the parametrized curve $\tilde{x}^{\mu}(\tau)\subset \tilde{M}$, which is the path followed by light traversing the transformation medium.  
Such direct visualization is one of the alluring features of transformation optics, in that the corporeal transformation can be elucidated from a deformation of an initial set of trajectories of light into a desired set of trajectories.
However, to take advantage of this procedure it is necessary to know the initial set of trajectories.
Since the parametrization of null geodesics on spacetime manifolds may be unfamiliar, here we explain how to obtain the parametrization of null geodesics in Friedmann-Lema\^{\i}tre-Robertson-Walker spacetime. 
We do this in some detail because a complete derivation is not readily found in the literature.

We parameterize a curve with affine parameter $\tau$, along with its velocity and acceleration as
\begin{subequations}
	\begin{equation} \label{Eq:AffineLocation}
	x^{\mu}(\tau)=\left(x^0(\tau),x^1(\tau),x^2(\tau),x^3(\tau)\right)
	\end{equation}
	\begin{equation} \label{Eq:AffineVelocity}
	v^{\mu}(\tau)= \frac{dx^{\mu}}{d\tau} = \dot{x}^{\mu} = \left(\dot{x}^0(\tau),\dot{x}^1(\tau),\dot{x}^2(\tau),\dot{x}^3(\tau)\right)
	\end{equation}
	\begin{equation} \label{Eq:AffineAcceleration}
	a^{\mu}(\tau) = \frac{d^2x^{\mu}}{d\tau^2} = \ddot{x}^{\mu} = \left(\ddot{x}^0(\tau),\ddot{x}^1(\tau),\ddot{x}^2(\tau),\ddot{x}^3(\tau)\right)
	\end{equation}
\end{subequations}
where
\begin{equation}
\dot{x}^0(\tau) = \frac{dx^0}{d\tau}, \quad \ddot{x^0}(\tau) = \frac{d^2 x^0}{d\tau^2}, \quad \text{etc.}
\end{equation}
It should be mentioned that although the point $x^{\mu}=(x^0,x^1,x^2,x^3)$ is often thought of as a ``vector,'' such a position vector is only a point on the manifold and is not an element of the tangent space, as $v^{\mu}$ is.
In this setting, $x^{\mu}$ is not a vector and does not transform like a vector.
Thus a curve $x^{\mu}(\tau)$ transforms with the map $T$ itself, while the vector field $v^{\mu}$ transforms via the pushforward of $T$ (assuming it exists).

\subsection{Null Geodesics in Minkowski Spacetime}
Before considering null geodesics in Friedmann Robertson Walker spacetimes, it is worthwhile to review the explicit affine parameterization of null geodesics in Minkowski spacetime, first in Cartesian coordinates and then in spherical coordinates.
\subsubsection{Cartesian coordinates}
In Cartesian coordinates the line element for Minkowski spacetime is
\begin{equation}
 ds^2= -c^2 dt^2 + dx^2 + dy^2 + dz^2.
\end{equation}
For a null geodesic we require
\begin{equation} \label{Eq:MinkowskiNullCondition}
 g_{\alpha\beta}v^{\alpha}v^{\beta} = -c^2\dot{t}^2+\dot{x}^2+\dot{y}^2+\dot{z}^2 = 0 .
\end{equation}
The geodesic equation is
\begin{equation}
 \ddot{x}^{\mu} + \Gamma^{\mu}_{\alpha\beta} \dot{x}^{\alpha}\dot{x}^{\beta},
\end{equation}
but in Cartesian coordinates $\Gamma^{\mu}_{\alpha\beta}=0$ so we are left with the set of equations $\ddot{x}^{\mu}=0$.
These integrate immediately to
\begin{equation}
\left(t(\tau),x(\tau),y(\tau),z(\tau)\right) = (u_t,u_x,u_y,u_z)\tau + (t_0,x_0,y_0,z_0)
\end{equation}
with the requirement that $u_x^2+u_y^2+u_z^2=u_t^2c^2$, and we generally set $c=1$.
\subsubsection{Spherical polar coordinates}
Integrating the geodesic equations in Cartesian coordinates was immediate, but things get messy in spherical polar coordinates. The Minkowskian line element is now
\begin{equation}
 ds^2 = -c^2 dt^2 + dr^2 +r^2 d\theta^2 + r^2\sin^2\theta\, d\varphi^2
\end{equation}
and the null condition is
\begin{equation}
  g_{\alpha\beta}v^{\alpha}v^{\beta} = -c^2\dot{t}^2+\dot{r}^2+r^2(\dot{\theta}^2+\sin^2\theta\dot{\varphi}^2) = 0 .
\end{equation}
The geodesic equations are now
\begin{subequations}
 \begin{equation}
  \ddot{t}=0
 \end{equation}
 \begin{equation}
  \ddot{r} - r(\dot{\theta}^2+\sin^2\theta\dot{\varphi}^2) = 0
 \end{equation}
 \begin{equation}
  \ddot{\theta} + 2\frac{\dot{r}}{r}\dot{\theta} - \cos\theta\sin\theta \dot{\varphi}^2 = 0
 \end{equation}
 \begin{equation}
  \ddot{\varphi} + 2\frac{\dot{r}}{r}\dot{\varphi} + 2\cot\theta \dot{\theta} \dot{\varphi} = 0.
 \end{equation}
\end{subequations}
The $t$ equation of course integrates immediately to the same result as previously obtained.
This should be expected since transforming between Cartesian and spherical coordinates leaves $t$ unchanged.
Although the $\varphi$ equation can readily be integrated to
\begin{equation}
 \dot{\varphi} = \omega_0\left(\frac{r_0\sin\theta_0}{r\sin\theta}\right)^2 
\end{equation}
we quickly run into difficulties finding an analytic solution to the $\theta$ equation.
Things can be simplified by restricting motion to the equatorial plane, in which case $\theta = \theta_0 = \pi/2$.
With this restriction,
 \begin{equation} \label{Eq:MinkPhiEq}
  \dot{\varphi} = \omega_0\left(\frac{r_0}{r}\right)^2
 \end{equation}
and the radial equation becomes
\begin{equation}
 \ddot{r} - \frac{\omega_0^2r_0^4}{r^3} = 0
\end{equation}
which, with 
\begin{equation}
 \dot{r}(\tau=0) = u_r,
\end{equation}
and the null condition
\begin{equation}
 u_r^2 + \omega_0^2 r_0^2 = u_t^2c^2.
\end{equation}
can be directly integrated to
\begin{equation}
 r(\tau) = \left[\left(\frac{r_0^2 \omega_0}{u_tc}\right)^2+\left(r_0\frac{u_r}{u_tc}\pm u_tc\tau\right)^2\right]^{1/2}.
\end{equation}
It is easy to verify that for purely radial geodesics, in which case $\omega_0=0$ and $u_r=c$, $r(\tau)=r_0\pm c\tau$ as expected.
Knowing $r(\tau)$, Eq.~(\ref{Eq:MinkPhiEq}) can now be integrated to find the parametrization of $\varphi$
\begin{equation}
 \varphi(\tau) = \pm\tan^{-1}\left[\frac{r_0 u_r \pm u_t^2c^2\tau}{r_0^2\omega_0}\right] \mp \tan^{-1}\left[\frac{u_r}{r_0\omega_0}\right] + \varphi_0.
\end{equation}

Choosing an initial point $(t_0,r_0,\pi/2,\varphi_0)$ and initial velocity with respect to $\tau$ $(u_t,u_r,0,\omega_0)$ we then have an explicit expression for the curve.  
Furthermore, one can readily convert this to Cartesian coordinates with $x(\tau) = r(\tau)\cos(\varphi(\tau))$, and $y(\tau) = r(\tau)\sin(\varphi(\tau))$ if, for example, one wanted to make a parametric plot with something like Mathematica.  
These become
\begin{equation}
 \begin{gathered}
   x = (u_r \cos\varphi_0 - r_0\omega_0\sin\varphi_0)\tau + r_0\cos\varphi_0\\
   y = (u_r \sin\varphi_0 + r_0\omega_0\cos\varphi_0)\tau + r_0\sin\varphi_0.
 \end{gathered}
\end{equation}
Note that if the chosen initial point on the trajectory coincides with the point of closest approach, then $r_0=b$ is just the impact parameter, $u_r=0$ because at the point of closest approach the velocity is purely tangential, and $\varphi_0 = \varphi_{in}+\pi/2$ where $\varphi_{in}$ is the angle at which the trajectory crosses the x-axis (special care is needed for trajectories with initial point at the origin).  If one is unconcerned with following the wavefront and only cares about the geometrical path, then rescaling $\tau \to r_0\omega_0\tau$ allows us to write
\begin{equation}
 \begin{gathered}
  x = \tau\cos\varphi_{in} - b\sin\varphi_{in}\\
  y = \tau\sin\varphi_{in} + b\cos\varphi_{in}.
 \end{gathered}
\end{equation}

\subsection{Null geodesics in FLRW: comoving Cartesian coordinates}
\subsubsection{Null geodesics}
In comoving Cartesian coordinates the FLRW metric is
\begin{equation}
 ds^2=-c^2dt^2 + a^2(t)\left(dx^2+dy^2+dz^2\right).
\end{equation}
We parameterize a curve with affine parameter $\tau$, along with it's velocity and acceleration, in the same manner as Eqs.\ (\ref{Eq:AffineLocation})--(\ref{Eq:AffineAcceleration}).
It should be borne in mind that the scale factor $a(t)$ appearing in the metric, which is a scalar function, is different from the acceleration vector $a^{\mu}$.  
Note that with this parameterization, the scale factor is also parametrized by $\tau$ through its dependence on $t$, in other words $a=a(t(\tau))$.  
We will suppress the explicit dependence on $t$ or $\tau$ except where necessary for clarity.  Dot notation will denote derivatives with respect to $\tau$ and primes will denote derivatives with respect to $t$
\begin{equation}
 a' = a'(t) = \frac{da}{dt}, \quad \dot{a} = \frac{da}{d\tau} = a'\dot{t}. 
\end{equation}
For a null geodesic we require
\begin{equation} \label{Eq:NullCondition}
 g_{\alpha\beta}v^{\alpha}v^{\beta} = -c^2\dot{t}^2+a^2(\dot{x}^2+\dot{y}^2+\dot{z}^2) = 0.
\end{equation}
The geodesic equation yields
\begin{subequations}
 \begin{equation}
  \ddot{t} + \frac{a  a'}{c^2}(\dot{x}^2+\dot{y}^2+\dot{z}^2) = 0, \label{Eq:Geot}
 \end{equation}
 \begin{equation}
  \ddot{x} + \frac{2a'}{a}\dot{t}\dot{x} = 0. \label{Eq:Geox}
 \end{equation}
\end{subequations}
The equations for $y$ and $z$ are identical to that of $x$, thus it is sufficient to consider only the equations for $t$ and $x$. 
Using the null condition Eq.~(\ref{Eq:NullCondition}) and $a'\dot{t} = \dot{a}$, the $t$ and $x$ equations become
\begin{subequations}
\begin{equation} \label{Eq:tDiffEq}
\ddot{t} + \frac{\dot{a}}{a}\dot{t} = 0,
\end{equation}
\begin{equation}
\ddot{x} + 2\frac{\dot{a}}{a}\dot{x} = 0
\end{equation}
\end{subequations}
with solutions
\begin{subequations}
	\begin{equation} \label{Eq:tSoln}
	t(\tau) = u_t a_0\int\frac{1}{a} d\tau,
	\end{equation}
	\begin{equation}\label{Eq:xSoln}
	x(\tau) = u_x a_0^2\int\frac{1}{a^2}d\tau,
	\end{equation}
\end{subequations}
and similarly for $y$ and $z$.
The coefficients $(u_t,u_x,u_y,u_z)$ give the direction of propagation of the geodesic at $\tau=0$, i.e.\ $v^{\mu}|_{\tau=0}=(u_t,u_x,u_y,u_z)$, and satisfy the null condition
\begin{equation}
 -c^2u_t^2+a_0^2(u_x^2+u_y^2+u_z^2)=0.
\end{equation}
\subsubsection{Conformal flatness}
It is well known that FLRW is conformally flat, which means that FLRW is equivalent to a conformal rescaling of Minkowski space.  
In other words, transforming to ``conformal time'' 
\begin{equation}
 d\eta = \frac{1}{a(t)} dt,
\end{equation}
the metric becomes 
\begin{equation}
 ds^2= a(t)\left(-c^2d\eta^2 + dx^2+dy^2+dz^2\right).
\end{equation}
Since the null character of a tangent vector (and hence the null character of a geodesic) is invariant to conformal transformations, null geodesics of Minkowski spacetime are conformally mapped to null geodesics of FLRW spacetime.  
In Minkowski spacetime, a defining characteristic of null geodesics is that $\Delta x = c\, \Delta t$ for light propagating along the $x$-axis, and this relation between intervals should also be reflected in FLRW if we switch to conformal time.  
Indeed, transforming Eq.~(\ref{Eq:tDiffEq}) to conformal time, we find the equation
\begin{equation}
 \ddot{\eta} + 2\frac{\dot{a}}{a}\dot{\eta} = 0
\end{equation}
with corresponding solution
\begin{equation}
  \eta(\tau) = u_\tau a_0^2\int\frac{1}{a^2}d\tau.
\end{equation}
Thus the intervals $\Delta x = c\, \Delta\eta$ clearly show the conformal flatness of FLRW.

\subsection{Evaluating the scale factor}
To proceed further requires knowledge about the form of the scale factor function $a(\tau)$.  The scale factor cannot be specified arbitrarily, but instead must be obtained from Einstein's equations, $G_{\mu\nu}=8\pi T_{\mu\nu}$. 
It is known that the FLRW metric is the solution for a spacetime filled with a perfect fluid.  
The stress tensor $T_{\mu\nu}$ for a perfect fluid with 4-velocity $u^{\mu}$ is given by
\begin{equation}
 T_{\mu\nu} = \rho u_{\mu}u_{\nu} + \frac{p}{c^2}(g_{\mu\nu}+u_{\mu}u_{\nu})
\end{equation}
where $\rho$ is the energy density and $p$ is the pressure.
Raising one index, $T\indices{_{\mu}^{\nu}}$ has matrix form
\begin{equation}
 T\indices{_{\mu}^{\nu}} = \begin{pmatrix}
  -\rho & 0 & 0 & 0\\
  0 & \frac{p}{c^2} & 0 & 0\\
  0 & 0 & \frac{p}{c^2} & 0\\
  0 & 0 & 0 & \frac{p}{c^2}
 \end{pmatrix}.
\end{equation}
The continuity equation 
\begin{equation} \label{Eq:FlatContinuity}
 \rho'+3\frac{a'}{a}\left(\rho+\frac{p}{c^2}\right) = 0
\end{equation}
may be obtained from the divergence-free condition on the stress tensor $T\indices{_{\mu}^{\nu}_{;\nu}}=0$.
Letting $w=p/c^2\rho$ be the \textit{equation of state} and dividing through by $\rho$, Eq.~(\ref{Eq:FlatContinuity}) is rewritten as
\begin{equation}
 \frac{\rho'}{\rho} + 3(1+w)\frac{a'}{a}= 0.
\end{equation}
For the perfect fluids upon which most cosmological models are based, the equation of state is a constant, and we therefore find
\begin{equation}
 \rho = \rho_0 \left(\frac{a_0}{a}\right)^{3(1+w)}.
\end{equation}

Next, there are two independent Einstein equations
\begin{subequations}
 \begin{equation} \label{Eq:CartesianComovingEinstein1}
  \left(\frac{a'}{ca}\right)^2 = \frac83\pi\rho,
 \end{equation}
and
 \begin{equation} \label{Eq:CartesianComovingEinstein2}
  2\frac{a''}{c^2a} +\left(\frac{a'}{ca}\right)^2 = -8\pi \frac{p}{c^2}.
 \end{equation}
\end{subequations}
In the previous subsection on Cartesian coordinates we found the relations $\dot{a}=\dot{t}a'$ and $\dot{t}=u_t a_0 a^{-1}$.  
Given that changing to comoving spherical polar coordinates, or even physical spherical coordinates, leaves the time coordinate invariant we should expect the equation for $t(\tau)$ to remain unchanged, as will be explicitly demonstrated below. 
Hence $a'=\tfrac{a\dot{a}}{u_ta_0}$ in all coordinates under consideration, and Eq.~(\ref{Eq:CartesianComovingEinstein1}) becomes
\begin{equation}
 \frac{\dot{a}^2}{u_t^2 c^2 a_0^2} - \frac83 \pi\rho_0 \left(\frac{a_0}{a}\right)^{3(1+w)} = 0,
\end{equation}
which has solution
\begin{equation} \label{Eq:aSoln}
 a =  a_0\left[1 \pm \sqrt{\frac83 \pi \rho_0}\left(\frac{5 + 3 w}{2}\right)u_tc\tau\right]^{\frac{2}{5 + 3 w}}.
\end{equation}
Note that this solution looks rather different from the usual derivation of this sort, in particular, for a cosmological constant with $w=-1$ one normally obtains $a = a_0e^{Ht}$.
The difference stems from the fact that we are parameterizing with respect to the affine parameter $\tau$ rather than the coordinate time or, as is often done, the conformal time $\eta$.

The Cartesian parametrizations of null geodesics in FLRW is finally obtained by using the form of $a(\tau)$ given by Eq.~(\ref{Eq:aSoln}) in Eqs.~(\ref{Eq:tSoln}) and (\ref{Eq:xSoln}).
Choosing the positive branch we find, for $w\neq -1$
\begin{equation}
 t = \frac{2}{3(1+w)c}\left(\frac83 \pi \rho_0\right)^{-1/2} \left[\left(1 + \sqrt{\frac83 \pi \rho_0}\left(\frac{5 + 3 w}{2}\right)u_t c\tau\right)^{\frac{3(1+w)}{5 + 3 w}} -1\right] + t_0
\end{equation}
and
\begin{equation}
 x = \left(\frac{u_x}{u_t c}\right)\left(\frac{2}{1+3w}\right)\left(\frac83 \pi \rho_0\right)^{-1/2} \left[\left(1 + \sqrt{\frac83 \pi \rho_0}\left(\frac{5 + 3 w}{2}\right)u_t c\tau\right)^{\frac{1+3w}{5 + 3 w}} -1\right] + x_0,
\end{equation}
with similar results for $y$ and $z$.
For $w=-1$, $a(\tau)$ becomes
\begin{equation}
 a= a_0\left(1+\sqrt{\frac83\pi\rho_0}\ u_t c \tau\right)
\end{equation}
with
\begin{equation} \label{Eq:tComovingSoln}
 t = \frac1c\left(\frac83 \pi \rho_0\right)^{-1/2} \ln \left[1 \pm \sqrt{\frac83\pi\rho_0}\ u_t c \tau\right] + t_0
\end{equation}
and
\begin{equation} \label{Eq:DampedPeculiar}
 x = \left(\frac{u_x}{u_t c}\right)\left(\frac83 \pi \rho_0\right)^{-1/2}\left(1-\frac{1}{1+\tau\sqrt{\frac83 \pi \rho_0}}\right) + x_0. 
\end{equation}
One of the fascinating things about FLRW spacetimes is that the peculiar motion (that portion of the motion that is not due to the Hubble flow) of timelike objects gets damped out, and eventually the physical motion is due solely to the Hubble flow.  
In other words, everything eventually becomes comoving.  
But the explicit parametric equation for the $x$-coordinate of null geodesics Eq.\ (\ref{Eq:DampedPeculiar}) shows that the same holds true for light in a universe dominated by a cosmological constant, $w=-1$.  
As $\tau\to\infty$ the null geodesic limits to comoving coordinate
\begin{equation}
 x(\tau\to\infty)=x_0 + \left(\frac{u_x}{u_t c}\right)\left(\frac83 \pi \rho_0\right)^{-1/2}.
\end{equation}
The physical distance, of course, continues to increase such that a physical observer always measures $c=1$ for the speed of light.
Note that the scale factor when the light was emitted, $a_0$, enters $u_t c$ through the null condition. This shows that in a cosmological constant dominated universe light is limited in the comoving distance it is able to travel.  Thus all observers separated by a comoving distance
\begin{equation}
 D > \frac{1}{a_0}\left(\frac83 \pi \rho_0\right)^{-1/2}
\end{equation}
at $\tau=0$ are not causally connected.

\subsection{Null geodesics in FLRW: comoving spherical coordinates} \label{App:FRWComoving}
Next, let's consider spherical polar coordinates with line element given by Eq.\ (\ref{Eq:FRWComoving}).
The null condition is
\begin{equation} \label{Eq:SphericalNullCondition}
 -c^2\dot{t}^2 + a^2\left(\dot{r}_c^2 + r_c^2(\dot{\theta}^2+\sin^2\theta\dot{\varphi}^2)\right) = 0.
\end{equation}
and the geodesic equations are now
\begin{subequations}
 \begin{equation} \label{Eq:SpGeot}
  \ddot{t} + \frac{a a'}{c^2}\left(\dot{r}_c^2 + r_c^2(\dot{\theta}^2+\sin^2\theta\dot{\varphi}^2)\right) = 0
 \end{equation}
 \begin{equation} \label{Eq:SpGeor}
  \ddot{r}_c  + 2\frac{\dot{a}}{a}\dot{r}_c - r_c(\dot{\theta}^2+\sin^2\theta\dot{\varphi}^2) = 0.
 \end{equation}
 \begin{equation} \label{Eq:SpGeoTheta}
  \ddot{\theta} + 2\left(\frac{\dot{r}_c}{r_c}+\frac{\dot{a}}{a}\right)\dot{\theta}-\cos\theta\sin\theta\dot{\varphi}^2 = 0
 \end{equation}
 \begin{equation} \label{Eq:SpGeoPhi}
  \ddot{\varphi} + 2 \left(\frac{\dot{r}_c}{r_c}+\frac{\dot{a}}{a}+\cot\theta \dot{\theta}\right)\dot{\varphi} = 0
 \end{equation}
\end{subequations}
From the null condition and $\dot{a} = a'\dot{t}$, Eq.~(\ref{Eq:SpGeot}) leads immediately to
\begin{equation}
 \ddot{t}+\frac{\dot{a}}{a}\dot{t}=0.
\end{equation}
That the $t$ equation is the same as that found previously is entirely expected; the change of coordinate system from Cartesian to spherical leaves the $t$ coordinate unchanged, and hence the parametric equation for $t$ should also remain unchanged.
A first integral for $\varphi$ can also be found immediately
\begin{equation} \label{Eq:FRWPhiDot}
 \dot{\varphi} = \omega_0\left(\frac{r_{c_0}a_0\sin\theta_0}{r_ca\sin\theta}\right)^2.
\end{equation}
However, to find complete solutions it is advantageous to assume that $\theta=\pi/2$, in which case Eq.~(\ref{Eq:SpGeoTheta}) also yields a first integral
\begin{equation}
 \dot{\theta}=\dot{\theta}_0\left(\frac{r_{c_0}a_0}{r_ca}\right)^2.
\end{equation}
What this shows is that if we confine ourselves to the equatorial plane with $\dot{\theta}_0=0$, then we will remain in the equatorial plane.  
Finding the solution to the $r$ equation Eq.~(\ref{Eq:SpGeor}) is more difficult.  
In the limited case of purely radial null geodesics then $\dot{\theta}=\dot{\varphi}=0$, and the equation reduces to
\begin{equation}
 \frac{\ddot{r}_c}{\dot{r}_c}+2\frac{\dot{a}}{a}=0,
\end{equation}
which is straightforward to solve.

The task is somewhat less straightforward for non-radial null geodesics. As in the previous Minkowski example, we restrict to $\theta=\pi/2$, whence the radial equation can be written
 \begin{equation} 
  \frac{\ddot{r}_c}{\dot{r}_c} + 2\frac{\dot{a}}{a} - \frac{\omega_0^2r_0^4a_0^4}{\dot{r}r^3a^4} = 0.
 \end{equation}
The complexity of the last term prevents us from finding an analytic solution in terms of $r_c$ and $a$.
Although Eq.~(\ref{Eq:aSoln}) provides an explicit expression for $a(\tau)$, event this is too complicated to solve.
On the other hand, this differential equation for $r_c$ is second order, but the null condition gives us a first order equation
\begin{equation} \label{Eq:FRWrNull}
 a^2\dot{r}_c^2 + \frac{\omega_0^2r_{c_0}^4a_0^4}{a^2r_c^2} - \frac{a_0^2}{a^2} = 0.
\end{equation}
By substituting Eq.~(\ref{Eq:aSoln}) for $a(\tau)$, one eventually finds the solution
\begin{equation} \label{Eq:FRWrSoln}
 r_c(\tau) = \left[u_t^2c^2\left[\frac{2}{u_t c a_0(1+3w)}\left(\frac83\pi\rho_0\right)^{-\frac12} \left(\left(1\pm\sqrt{\frac83\pi\rho_0}\left(\frac{5+3w}{2}\right)u_t c \tau\right)^{\frac{1+3w}{5+3w}} - 1\right) \pm \frac{a_0r_{c_0}u_{r_c}}{u_t^2c^2}\right]^2 + \left(\frac{a_0r_{c_0}^2\omega_0}{u_tc}\right)^2\right]^{\frac12}.
\end{equation}
This can also be written in terms of $a(\tau)$ as
\begin{equation} \label{Eq:rComovingSoln}
 r_c(\tau) = \left[u_t^2 c^2\left[ \frac{2}{1+3w}\left( \frac{a_0}{a(\tau)\dot{a}(\tau)}-\frac{1}{\dot{a}_0}\right) \pm \frac{a_0r_{c_0}u_{r_c}}{u_t^2c^2}\right]^2 + \left(\frac{a_0r_{c_0}^2\omega_0}{u_tc}\right)^2\right]^{\frac12}
\end{equation}

Knowing $r_c(\tau)$ explicitly, Eq.~(\ref{Eq:FRWPhiDot}) can now be integrated to find an explicit parameterization of $\varphi(\tau)$
\begin{multline} \label{Eq:phiComovingSoln}
 \varphi(\tau) = \tan^{-1}\left[ \frac{c^2 u_t^2}{a_0 r_{c_0}^2\omega_0} \left(\frac{2}{u_t c a_0(1+3w)}\left(\frac83\pi\rho_0\right)^{-\frac12} \left(\left(1\pm\sqrt{\frac83\pi\rho_0}\left(\frac{5+3w}{2}\right)u_t c \tau\right)^{\frac{1+3w}{5+3w}} - 1\right) \pm \frac{a_0r_{c_0}u_{r_c}}{u_t^2c^2} \right) \right] \\ \mp \tan^{-1}\left[\frac{u_{r_c}}{r_{c_0}\omega_0}\right]+\varphi_0.
\end{multline}
Choosing an initial point $(t_0,r_{c_0},\pi/2,\varphi_0)$ and initial velocity $(u_t,u_{r_c},0,\omega_0)$ we then have an explicit parameterization for the curve.  Furthermore, one can readily convert this to Cartesian coordinates with $x_c(\tau)=r_c(\tau)\cos(\varphi(\tau))$ and $y_c(\tau)=r_c(\tau)\sin(\varphi(\tau))$.  These become
\begin{equation} \label{Eq:ComovingSphericaltoCartesianSoln}
 \begin{gathered}
 x_c(\tau) = \frac{2}{u_t c a_0(1+3w)}\left(\frac83\pi\rho_0\right)^{-\frac12} \left(\left(1\pm\sqrt{\frac83\pi\rho_0}\left(\frac{5+3w}{2}\right)u_t c \tau\right)^{\frac{1+3w}{5+3w}} - 1\right) \left(u_{r_c}\cos\varphi_0 - r_{c_0}\omega_0\sin\varphi_0 \right) + r_{c_0}\cos\varphi_0 \\
 y_c(\tau) = \frac{2}{u_t c a_0(1+3w)}\left(\frac83\pi\rho_0\right)^{-\frac12} \left(\left(1\pm\sqrt{\frac83\pi\rho_0}\left(\frac{5+3w}{2}\right)u_t c \tau\right)^{\frac{1+3w}{5+3w}} - 1\right) \left(u_{r_c}\sin\varphi_0 + r_{c_0}\omega_0\cos\varphi_0 \right) + r_{c_0}\sin\varphi_0.
 \end{gathered}
\end{equation}

It can be useful to write down these main results in a more compact notation.  We let 
\begin{equation}
 d=\frac{2}{5+3w} \quad \text{and} \quad A = u_tc\sqrt{\frac83\pi\rho_0}.
\end{equation}
Then we have
\begin{equation}
 a =  a_0\left[1 \pm \frac{A}{d}\tau\right]^{d}
\end{equation}
for the scale factor.  Equations (\ref{Eq:tComovingSoln}) and (\ref{Eq:DampedPeculiar}) become
\begin{equation}
 t = \frac{u_t d}{(1-d)A} \left[\left(1 + \frac{A}{d}\tau\right)^{1-d} -1\right] + t_0
\end{equation}
and
\begin{equation}
 x_c = \frac{u_xd}{(1-2d)A} \left[\left(1 + \frac{A}{d}\tau\right)^{1-2d} -1\right] + x_0.
\end{equation}
The spherical coordinate results Eqs.\ (\ref{Eq:rComovingSoln}) and (\ref{Eq:phiComovingSoln}) become
\begin{equation} \label{Eq:FRWrSoln2}
 r_c(\tau) = \left[u_t^2c^2\left[\frac{d}{a_0(1-2d)A} \left(\left(1\pm\frac{A}{d}\tau\right)^{1-2d} - 1\right) \pm \frac{a_0r_{c_0}u_{r_c}}{u_t^2c^2}\right]^2 + \left(\frac{a_0r_{c_0}^2\omega_0}{u_tc}\right)^2\right]^{\frac12}
\end{equation}
and
\begin{equation}
 \varphi(\tau) = \tan^{-1}\left[ \frac{c^2 u_t^2}{a_0 r_{c_0}^2\omega_0} \left(\frac{d}{a_0(1-2d)A} \left(\left(1\pm \frac{A}{d} \tau\right)^{1-2d} - 1\right) \pm \frac{a_0r_{c_0}u_{r_c}}{u_t^2c^2} \right) \right] \\ \mp \tan^{-1}\left[\frac{u_r}{r_{c_0}\omega_0}\right]+\varphi_0.
\end{equation}
While Eqs. (\ref{Eq:ComovingSphericaltoCartesianSoln}) become
\begin{equation}
 \begin{gathered}
 x_c(\tau) = \frac{d}{a_0(1-2d)A} \left(\left(1\pm\frac{A}{d} \tau\right)^{1-2d} - 1\right) \left(u_{r_c}\cos\varphi_0 - r_{c_0}\omega_0\sin\varphi_0 \right) + r_{c_0}\cos\varphi_0 \\
 y_c(\tau) = \frac{d}{a_0(1-2d)A} \left(\left(1\pm\frac{A}{d}\tau\right)^{1-2d} - 1\right) \left(u_{r_c}\sin\varphi_0 + r_{c_0}\omega_0\cos\varphi_0 \right) + r_{c_0}\sin\varphi_0.
 \end{gathered}
\end{equation}

As discussed in the Minkowski example, if one is only interested in is the shape of the curve and doesn't care whether the parameter is affine or not, then by rescaling with
\begin{equation}
 s= \frac{d}{a_0(1-2d)A} \left(\left(1\pm\frac{A}{d} \tau\right)^{1-2d} - 1\right)
\end{equation}
we recover
\begin{equation}
 \begin{gathered}
 x_c(\tau) = s \left(u_{r_c}\cos\varphi_0 - r_{c_0}\omega_0\sin\varphi_0 \right) + r_{c_0}\cos\varphi_0 \\
 y_c(\tau) = s \left(u_{r_c}\sin\varphi_0 + r_{c_0}\omega_0\cos\varphi_0 \right) + r_{c_0}\sin\varphi_0,
 \end{gathered}
\end{equation}
which is exactly the same as the spatial parametrizations in Minkowski spacetime.
With an additional rescaling of $t$ we have explicitly recovered the conformal flatness of FLRW in comoving spherical coordinates.  This may seem like a very roundabout way to get to a result that could have been found more or less immediately from conformal arguments,
however, the point of this approach is that we can find an explicit \textit{affine} parametrization of the geodesics in these coordinates, and the fact that we explicitly recover the conformality of the solutions is an important check on the calculations.

\subsection{Null geodesics in FLRW: physical coordinates} \label{App:FRWPhysical}
It is also advantageous to work in ``physical coordinates'' centered on the cloak, rather than co-moving coordinates.
The reason for this is that for the cloak to remain of fixed size, the individual elements making up the cloak must overcome the Hubble flow by whatever internal forces that are keeping the cloak together.  
This means that although we may consider the center of the cloak to be comoving, other points on the cloak are not comoving.
Therefore, a cloak of fixed physical size must actually be shrinking when described in comoving coordinates.  
Despite the fact that any coordinate system should be equally adequate for describing null geodesics, physical coordinates are may be more advantageous for describing a cloaking device of fixed physical size, and for describing the cloak parameters in a system in which all parts of the cloak are seen to be at rest with respect to the observer. 

Physical coordinates are reflective of the distances actually measured by an observer.  
A fixed comoving observer measures the physical distance from himself (at $r_p=0$) to a galaxy at fixed comoving coordinate $r_p$ to be $d=a(t)r_p$.  
We let this increasing physical distance be the physical radial coordinate, so $r_p = a(t)r$, where $r$ refers to the comoving coordinates.
From this relation we have
\begin{equation}
 dr_p = a'r_c dt + a dr_c = \frac{a'}{a}r_p dt + a dr_c,
\end{equation}
and therefore
\begin{equation}
 dr_c = \frac1a dr_p - \frac{a'}{a^2} r_p dt.
\end{equation}
Plugging this into the FLRW line element Eq.~(\ref{Eq:FRWComoving}), there results
\begin{equation} 
ds^2 = -\left(c^2 - \frac{\dot{a}(t)^2}{a(t)^2} r_p^2\right) dt^2 - 2\frac{\dot{a}(t)}{a(t)}r_p dr_p dt +  dr_p^2 + r_p^2 d\Omega^2
\end{equation}
Since the transformation from comoving to physical coordinates only involves transforming the $r$ coordinate, it is expected that the equations for $t$ and $\dot{\varphi}$ should remain essentially unchanged (we again assume equatorial motion).  Calculating the geodesic equations confirms that expectation, with
\begin{equation}
 \begin{gathered}
  \dot{t} =  \frac{u_t a_0}{a}, \\
  \dot{\varphi} = \omega_0\left(\frac{r_{p_0}}{r_p}\right)^2.
 \end{gathered}
\end{equation}
With motion constrained to the equatorial plane and these expressions for $\dot{t}$ and $\dot{\varphi}$ , the null condition becomes
\begin{equation}
 \dot{r}_p^2 - 2\frac{\dot{a}\dot{r}_pr_p}{a} + \frac{\dot{a}^2r_p^2}{a^2} + \frac{\omega_0^2r_{p_0}^4}{r_p^2} - \frac{u_t^2c^2 a_0^2}{a^2} = 0.
\end{equation}
But it is straightforward to show that this same equation may be obtained by starting with the null condition Eq.~(\ref{Eq:FRWrNull}) in comoving coordinates and making the replacement $r_c\to r_p/a$.
Therefore the solution may be found pretty much immediately from Eq.~(\ref{Eq:FRWrSoln}) or Eq.~(\ref{Eq:FRWrSoln2}), it is
\begin{equation}
 r_p(\tau) = \left(1+\frac{A}{d}\tau\right)^d\sqrt{u_t^2c^2\left(\frac{d}{(1-2d)A}\left(\left(1+\frac{A}{d}\tau\right)^{1-2d}-1\right)+\frac{r_{p_0}(u_{r_p}-Ar_{p_0})}{u_t^2c^2}\right)^2+\frac{r_{p_0}^4\omega_0^2}{u_t^2c^2} }.
\end{equation}
The equation for $\dot{\varphi}$ can now be directly integrated to find
\begin{equation}
 \varphi(\tau) = \tan^{-1}\left[\frac{u_t^2c^2}{r_{p_0}^2\omega_0}\left(\frac{d}{(1-2d)A} \left(\left( 1-\frac{A}{d}\tau \right)^{1-2d} -1\right) + \frac{r_{p_0}(u_{r_p}-Ar_{p_0}}{u_t^2c^2}  \right) \right] - \tan^{-1}\left[\frac{u_{r_p}-Ar_{p_0}}{r_{p_0}\omega_0} \right] + \varphi_0.
\end{equation}
Once again the Cartesian coordinates can be expressed as $x_p = r_p\cos\varphi$ and $y_p = r_p\sin\varphi$.  We find
\begin{equation}
\begin{gathered}
 x_p(\tau) = \frac{d}{(1-2d)A}\left(1+\frac{A}{d}\tau\right)^{2d}\left(\left(1+\frac{A}{d}\tau\right)^{1-2d}-1\right)\left((u_{r_p}-Ar_{p_0})\cos\varphi_0 - r_{p_0}\omega_0\sin\varphi_0\right) + \left(1+\frac{A}{d}\tau\right)^{2d} r_{p_0}\cos\varphi_0 \\
 y_p(\tau) = \frac{d}{(1-2d)A}\left(1+\frac{A}{d}\tau\right)^{2d}\left(\left(1+\frac{A}{d}\tau\right)^{1-2d}-1\right)\left((u_{r_p}-Ar_{p_0})\sin\varphi_0 + r_{p_0}\omega_0\cos\varphi_0\right) + \left(1+\frac{A}{d}\tau\right)^{2d} r_{p_0}\sin\varphi_0
\end{gathered}
\end{equation}

This is an interesting result that is substantially different to the usual behaviour an observer in Minkowski spacetime expects.  
As before, suppose that at $\tau=0$ the curve makes its closest approach to $r_p=0$, in which case the initial tangent to the curve has no radial component, $u_{r_p}=0$, and we can relate $\varphi_0$ to the angle $\varphi_{in}$ at which the trajectory crosses the x-axis via $\varphi_0=\varphi_{in}+\pi/2$.  
In both the Minkowski and comoving cases this allowed us to make the simplification
\begin{equation}
\begin{gathered}
 x(s) = s\cos\varphi_{in} - b\sin\varphi_{in}\\
 y(s) = s\sin\varphi_{in} + b\cos\varphi_{in},
\end{gathered}
\end{equation}
which carries with it the implication that if $\varphi_{in} = 0$ then the trajectory remains at fixed $y$ and if $\varphi_{in}=\pi/2$ the trajectory remains at fixed $x$.
But in the case of physical coordinates this is no longer the case.
Indeed, for a trajectory with $u_r=0$ and $\varphi_{in}=0$, the $y$ component continues to evolve as
\begin{equation}
 y(\tau) = r_{p_0}\left(1+\frac{A}{d}\tau\right)^{2d}\left[1+ \frac{d}{1-2d}\left(\left(1+\frac{A}{d}\tau\right)^{1-2d}-1\right) \right].
\end{equation}
Thus according to a physical observer, expansion of a congruence of null geodesics is inevitable and unavoidable.


\section{Coordinate transformations between comoving and physical coordinates} \label{App:CoordTrans}

The coordinate transformation between comoving coordinates $(t,r_c,\theta,\varphi)$ and physical coordinates $(t,r_p,\theta,\varphi)$ is given by
\begin{equation}
 r_p = a(t) r_c
\end{equation}
while the coordinate transformation from comoving to physical coordinates is given by the inverse of this, or
\begin{equation}
 r_c = \frac{1}{a(t)}r_p.
\end{equation}
We may make manifest the invariance of a tensorial object under this kind of diffeomorphism of the manifold to itself, call it $\mathcal{C}: M\to M$ and it's inverse $\mathcal{C}^{-1}$.  
The associated differential $d\mathcal{C}:T_{q_c}M\to T_{q_p}M$ defines a map from the tangent space at a point $x_c=q_c$ in comoving coordinates to the tangent space at the related point $x_p=q_p=\mathcal{T}(x_c)$ in physical coordinates.
The vectorial components $v_{q_c}^{\mu}$ at point $x_c = q$ in comoving coordinates transform to the vectorial components $v_{q_p}^{\nu'}$ (where primed indices refer to physical coordinates) at point $x_p = q_p = \mathcal{C}(p)$ in physical coordinates via the pushforward map given by the differential $d\mathcal{C}$.
This takes the form
\begin{equation}
 v_{q_p}^{\nu'} = \left. J^{\nu'}_{\mu}\right|_{q_c} v_{q_c}^{\mu} 
\end{equation}
where $J^{\nu'}_{\mu}$ is the Jacobian matrix of $\mathcal{C}$, evaluated at $q_c$, and is the matrix representation of the differential $d\mathcal{C}$.
Note that if $J^{\nu'}_{\mu}$ depends explicitly on the coordinate function $x_c$, then the components of the vector field $v^{\nu'}$ will be expressed in terms of $x_c$, so we must furthermore use the coordinate transformation $\mathcal{C}$ to transform the coordinate function appearing in $v^{\nu'}$.

Next consider the mapping of differential forms.  A 1-form $\omega_{\mu}$ in comoving coordinates must be pulled back to physical coordinates via the pullback map associated to $\mathcal{C}^{-1}$ rather than $\mathcal{C}$ itself.  The covectorial components $\omega_{\mu}(q_c)$ at point $q_c$ in comoving coordinates transform to covectorial components $\omega^{\nu'}(q_p)$ (where primed indices refer to physical coordinates) at point $x_p = q_p = \mathcal{C}(p)$ in physical coordinates via the pullback map given by the differential $d\mathcal{C}^{-1}$.
This takes the form
\begin{equation}
 \omega_{\nu'}(q_c) = \left. L^{\mu}_{\nu'}\right|_{q_p} \omega_{\mu}(q_c) 
\end{equation}
where $L^{\mu}_{\nu'}$ is the Jacobian matrix of $\mathcal{C}^{-1}$ and is evaluated at the point $q_p$.
Since $\mathcal{C}^{-1}$ is the inverse of $\mathcal{C}$, the two Jacobian matrices are related.  In particular
\begin{equation}
 L^{\mu}_{\nu'} = \left.(J^{\nu'}_{\mu})^{-1}\right|_{x_c\to \mathcal{C}(x_c)},
\end{equation}
and the coordinate invariance of a tensorial object follows
\begin{equation} \label{Eq:CoordInvariance}
 \mathbf{v}(x_c) = \left.\left.\left. v_{q_c}^{\mu} J^{\nu'}_{\mu}\right|_{q_c} L^{\mu}_{\nu'}\right|_{q_p}\frac{\partial}{\partial x_c^{\mu}} \right|_{x_c\to \mathcal{C}(x_c)} =  v_{q_p}^{\nu'} \frac{\partial}{\partial x_p^{\nu'}} = \mathbf{v}(x_p).
\end{equation}
Now, if we want to transform the tensor components $\chi\indices{_{\alpha\beta}^{\mu\nu}}(x_c)$ from comoving to physical coordinates, we apply
\begin{equation} \label{Eq:ChiComovingToPhysical}
 \chi\indices{_{\alpha'\beta'}^{\mu'\nu'}}(x_p) =  \left. L^{\alpha}_{\alpha'}L^{\beta}_{\beta'} \chi\indices{_{\alpha\beta}^{\mu\nu}}(x_c) J^{\mu'}_{\mu}J^{\nu'}_{\nu} \right|_{x_c\to \mathcal{C}(x_c)}.
\end{equation}

A similar construction allows us to transform from physical coordinates back to comoving coordinates.  
The essential difference between a coordinate and corporeal transformation is that a coordinate transformation transforms both the components and the basis elements, as in Eq.\ (\ref{Eq:CoordInvariance}), whereas a corporeal transformation acts only upon the components, while leaving the basis elements intact.

\section{Local frame transformations} \label{App:FrameTrans}
In curved spacetimes we often deal with coordinate patches where the coordinates are not orthonormal.  
This implies that the metric appears as something other than the Minkowski metric.  
It is desirable to be able to construct a \textit{locally flat frame}, relative to which a vector, covector, or mixed tensor may be described.

Begin with vectors.  For a given choice of coordinates there exists a set of associated vector fields $\{\frac{\partial}{\partial x^{\alpha}}\}$.  This set of basis vector fields span the tangent bundle over some portion of the manifold. Relative to this set of coordinate basis vector fields, any vector field may be decomposed as
\begin{equation}
 \mathbf{V} = V^{\alpha}\frac{\partial}{\partial x^{\alpha}}.
\end{equation}
These basis vector fields are chosen out of convenience because they have a natural association to the chosen coordinates.  
However, even with the chosen coordinates it is possible to choose a different set of basis vector fields that will span the tangent bundle over the same portion of the manifold.  
Let us label this new set of basis vector fields $\{e_A\}$, which is often called a tetrad or vierbein field.
Since the basis sets $\{\frac{\partial}{\partial x^{\alpha}}\}$ and $\{e_A\}$ both span the same vector (tangent bundle) space, we may express one set as a linear combination of the other,
\begin{equation}
 e_A = e\indices{_A^{\alpha}} \frac{\partial}{\partial x^{\alpha}}.
\end{equation}
Generally, the set $\{e_A\}$ is not associated to a coordinate system in the same way that the set $\frac{\partial}{\partial x^{\alpha}}$ is.  
We therefore refer to the set $\{e_A\}$ as a \textit{non-coordinate} set of basis vector fields.  
Since they both span the tangent bundle, the set $\{e_A\}$ is just as good a choice as $\{\frac{\partial}{\partial x^{\alpha}}\}$ for decomposing a vector field.  Thus the vector field $\mathbf{V}$ that was previously decomposed over the coordinate basis vector fields may now be decomposed over the non-coordinate vector fields
\begin{equation}
 \mathbf{V} = V^A e_A.
\end{equation}
One can think of the usual coordinate basis vector fields as a specially chosen tetrad field, subject to conditions discussed in section \ref{Sec:Discussion}.

Similarly, associated to the chosen coordinate system we have the set of basis covector fields (1-forms) $\{dx^{\mu}\}$.  
This set of basis 1-forms spans the cotangent bundle over the same portion of the manifold as above.  Relative to this basis, any 1-form $\boldsymbol{\omega}$ may be decomposed as
\begin{equation}
 \boldsymbol{\omega} = \omega_{\alpha} dx^{\alpha}.
\end{equation}
Again, these basis covector fields are chosen out of convenience because they have a natural association to the chosen coordinates.  
However, even with the chosen coordinates it is possible to choose a different set of basis covector fields that span the cotangent bundle over the same portion of the manifold.  
In particular, we will choose a set of covector fields $\{\theta^{A}\}$ that are dual to the vector fields chosen above such that $\theta^A e_B = \delta^A_B$, just as for coordinate basis fields we have the dual relation $dx^{\mu}\frac{\partial}{\partial x^{\nu}} = \delta^{\mu}_{\nu}$.  
Since the basis sets $\{\theta^{A}\}$ and $\{dx^{\mu}\}$ both span the same vector (cotangent bundle) space, we may express one set as a linear combination of the other.  We let
\begin{equation}
 \theta^A = e\indices{^A_{\mu}} dx^{\mu}.
\end{equation}
Since the set $\{\theta^A\}$ is not associated to a coordinate system in the same way that $\{dx^{\mu}\}$ is, we refer to $\{\theta^A\}$ as a \textit{non-coordinate} set of basis covector fields.

Because of the dual nature of both the coordinate and non-coordinate sets of basis fields, we have
\begin{equation}
 \theta^A e_B = (e\indices{^A_{\mu}} dx^{\mu})(e\indices{_B^{\alpha}} \frac{\partial}{\partial x^{\alpha}}) = (e\indices{^A_{\mu}}e\indices{_B^{\alpha}}) dx^{\mu} \frac{\partial}{\partial x^{\alpha}} = (e\indices{^A_{\mu}}e\indices{_B^{\alpha}}) \delta^{\mu}_{\alpha} = e\indices{^A_{\mu}}e\indices{_B^{\mu}}.
\end{equation}
But since by duality $\theta^A e_B = \delta^A_B$ we have a condition on the coefficients, namely that
\begin{equation}
 e\indices{^A_{\mu}}e\indices{_B^{\mu}} = \delta^A_B.
\end{equation}

Now, since
\begin{equation}
 \mathbf{V} = V^A e_A = V^A  e\indices{_A^{\alpha}} \frac{\partial}{\partial x^{\alpha}} =  (V^A  e\indices{_A^{\alpha}}) \frac{\partial}{\partial x^{\alpha}} = V^{\alpha}\frac{\partial}{\partial x^{\alpha}}
\end{equation}
we have a way of transforming the coefficients of a vector field from the coordinate basis to the non-coordinate basis, and similarly for 1-forms.
Thus we know how to transform both the coefficients and basis elements that describe vector fields and 1-forms.
Transformation of mixed tensor fields follows straightforwardly in the usual manner.

Given a line element 2-form $ds^2$, it can be decomposed relative to either the coordinate basis fields, or any other set of basis fields, 
\begin{equation} \label{Eq:MetricFrameTrans}
 ds^2 = g_{AB}\theta^A\theta^B = e\indices{_A^{\alpha}}e\indices{_B^{\beta}}g_{\alpha\beta} e\indices{^A_{\alpha}}dx^{\alpha} e\indices{^B_{\beta}}dx^{\beta} = g_{\alpha\beta}dx^{\alpha}dx^{\beta}.
\end{equation}
In particular, we may choose a set of tetrad basis fields such that $g_{AB}$ is the Minkowski metric.  
However, that metric does not refer to Minkowski coordinates or even to the Minkowski coordinate basis covector fields.  Instead, it refers to the non-coordinate basis covector fields $\{\theta^A\}$.

To summarize, we may decompose a vector field $\mathbf{V}$, covector field $\boldsymbol{\omega}$ or mixed tensor $\boldsymbol{\chi}$ with respect to either coordinate or non-coordinate basis fields, and this can be done regardless of the choice of coordinates.  A tetrad comprises a chosen set of basis vector fields at each coordinate point, but they are independent of the coordinate basis at that point.

\section{Coordinate transformation between local frames} \label{App:CoordFrameTrans}
Given $\chi\indices{_{\alpha\beta}^{\mu\nu}}(x_c)$ in comoving coordinates, we now understand this symbol to represent the coefficients of the tensor object $\boldsymbol{\chi}$ decomposed relative to the sets of coordinate basis vector and covector fields (or some tensor product of them).
From the previous discussion of local frames and tetrads, we can instead decompose $\boldsymbol{\chi}$ relative to a chosen frame where the metric is Minkowskian
\begin{equation}
 \chi\indices{_{AB}^{CD}}(x_c) = e\indices{_A^{\alpha}} e\indices{_B^{\beta}} e\indices{^C_{\mu}} e\indices{^D_{\nu}} \chi\indices{_{\alpha\beta}^{\mu\nu}}(x_c).
\end{equation}
On the other hand, in physical coordinates we have $\chi\indices{_{\alpha'\beta'}^{\mu'\nu'}}(x_p)$ as the coefficients of the same object $\boldsymbol{\chi}$ decomposed relative to the coordinate basis vector and covector fields of physical coordinates.
But we can just as well describe this relative to a locally flat set of basis vector fields as
\begin{equation} \label{Eq:AppChiPhysicalToLocal}
 \chi\indices{_{A'B'}^{C'D'}}(x_p) = e\indices{_{A'}^{\alpha'}} e\indices{_{B'}^{\beta'}} e\indices{^{C'}_{\mu'}} e\indices{^{D'}_{\nu'}} \chi\indices{_{\alpha'\beta'}^{\mu'\nu'}}(x_p).
\end{equation}
We know how to transform the coordinate decomposition of $\boldsymbol{\chi}$ from comoving to physical coordinates, so the remaining question is how do we transform from the non-coordinate decomposition of $\boldsymbol{\chi}$ in a local frame of comoving coordinates to the non-coordinate decomposition of $\boldsymbol{\chi}$ in a local frame of physical coordinates?
From Eqs.\ (\ref{Eq:ChiComovingToPhysical}) and (\ref{Eq:AppChiPhysicalToLocal}) this should be
\begin{equation}
 \chi\indices{_{A'B'}^{C'D'}}(x_p) =  e\indices{_{A'}^{\alpha'}} e\indices{_{B'}^{\beta'}} e\indices{^{C'}_{\mu'}} e\indices{^{D'}_{\nu'}}
\left. L^{\alpha}_{\alpha'}L^{\beta}_{\beta'} e\indices{^A_{\alpha}}e\indices{^B_{\beta}}e\indices{_C^{\mu}}e\indices{_D^{\nu}} \chi\indices{_{AB}^{CD}}(x_c)  
 J^{\mu'}_{\mu}J^{\nu'}_{\nu} \right|_{x_c\to \mathcal{T}(x_c)}.
\end{equation}
Combining the frame transformations with the coordinate transformations leads to
\begin{equation}
 \chi\indices{_{A'B'}^{C'D'}}(x_p) =  K\indices{_{A'}^A} K\indices{_{B'}^B} M\indices{^{C'}_C}M\indices{^{D'}_D}  \left. \chi\indices{_{AB}^{CD}}(x_c)\right|_{x_c\to \mathcal{T}(x_c)}
\end{equation}
with
\begin{equation}
 K\indices{_{A'}^A} =  e\indices{^A_{\alpha}} L^{\alpha}_{\alpha'} e\indices{_{A'}^{\alpha'}}
\end{equation}
and
\begin{equation}
 M\indices{^{C'}_C} =  e\indices{_C^{\mu}}  J^{\mu'}_{\mu}  e\indices{^{C'}_{\mu'}}.
\end{equation}

\begin{acknowledgments}
R.T.\ would like to thank J\"{o}rg Frauendiener for useful discussions.  R.T.\ is supported by the Royal Society of New Zealand through Marsden Fund Fast Start grant UOO1219.
\end{acknowledgments}


%

\end{document}